\documentclass[journal]{IEEEtran}
\usepackage{amsmath,graphicx,amsfonts, mathtools}
\usepackage{xcolor}
\usepackage{float}
\usepackage{algorithm}%
\usepackage{bm}
\usepackage{caption}
\usepackage{subcaption}
\usepackage[noadjust]{cite}
\usepackage{algpseudocode}%
\usepackage{amssymb}
\usepackage{tikz,pgfplots,pgfplotstable} 
\pgfplotsset{compat=1.12}
\usepgfplotslibrary{groupplots,fillbetween,colorbrewer,statistics}
\usetikzlibrary{intersections,calc,arrows,matrix,spy,pgfplots.statistics, pgfplots.colorbrewer}
\pgfplotsset{plot coordinates/math parser=false}
\usepackage{color}
\definecolor{magenta}{rgb}{0.7,0,0.7}
\definecolor{darkred}{rgb}{0.7,0,0}
\definecolor{darkgreen}{rgb}{0,0.5,0}
\definecolor{darkblue}{rgb}{0,0,0.7}
\definecolor{SkyBlue}{rgb}{0.53, 0.81, 0.92}

\usepackage[colorlinks]{hyperref}

\newcommand{\argmax}{\operatornamewithlimits{argmax}}
\newcommand{\argmin}{\operatornamewithlimits{argmin}}

\title{An Adversarial Learning Based Approach for Unknown View Tomographic Reconstruction}

\author{Mona Zehni, Zhizhen Zhao
\thanks{Thanks to NSF DMS-1854791, NSF OAC-1934757, and Alfred P. Sloan Foundation for funding.}
 \thanks{M. Zehni and Z. Zhao are with the Department of Electrical and Computer Engineering and Coordinated Science Laboratory, University of Illinois at Urbana-Champaign, Urbana, IL 61801 USA (e-mail: mzehni2@illinois.edu;  zhizhenz@illinois.edu).}} 

\begin{document}
\maketitle
\begin{abstract}
The goal of 2D tomographic reconstruction is to recover an image given its projections from various views. It is often presumed that projection angles associated with the projections are known in advance. Under certain situations, however, these angles are known only approximately or are completely unknown. It becomes more challenging to reconstruct the image from a collection of random projections. 
We propose an adversarial learning based approach to recover the image and the projection angle distribution by matching the empirical distribution of the measurements with the generated data. Fitting the distributions is achieved through solving a min-max game between a generator and a critic based on Wasserstein generative adversarial network structure.  
To accommodate the update of the projection angle distribution through gradient back propagation, we approximate the loss using the Gumbel-Softmax reparameterization of samples from discrete distributions. 
Our theoretical analysis verifies the unique recovery of the image and the projection distribution up to a rotation and reflection upon convergence. Our extensive numerical experiments showcase the potential of our method to accurately recover the image and the projection angle distribution under noise contamination. 
\end{abstract}
\vspace{-5pt}

\vspace{-0.2cm}
\section{Introduction}
\label{sec:intro}
Multitude of imaging modalities rely on reconstructing an unknown signal either in 2D or 3D domain given a set of partial measurements. Examples of such are medical imaging, cryo-electron microscopy (cryo-EM) and optical microscopy. More specifically, in a tomographic reconstruction setup, the measurements i.e. projections, are the line or plane integrals of the underlying object along various angles. In imaging applications such as CT, the projection angles are known a-priori through the acquisition process. However, this does not hold if the underlying object is moving or when reconstructing macromolecular structures in cryo-EM. Thus, it is important to develop solutions for tomographic reconstruction with unknown projection angles.
In this paper, we focus on 2D unknown view tomography with the ultimate goal of recovering the unknown object given a large set of noisy projections. 

Tomographic inversion with known viewing angles is typically a linear inverse problem and is solved by filtered back-projection (FBP), direct Fourier methods~\cite{start1981}, or solving a regularized optimization problem~\cite{Sidky_2008, Niu_2014, zhang2018, Gong2019}. However, the knowledge of the projection angles is not always available or accurate. To avoid adverse effects on the quality of the reconstructed image, it is important to account for uncertainties in the projection angles. To address this, one family of solutions determine the projection angles first~\cite{Basu2000, Basu2000_1, Fang2010, Coifman2008, Singer2013, Phan2017} and then reconstruct the image given the estimated projection views. Other approaches include iterative methods that solve for the 2D image and the projection angles in alternating steps~\cite{Cheikh2017}. While proven effective, these methods are computationally expensive and sensitive to initialization.
In another class of methods, to circumvent the estimation/refinement of the projection angles, a set of rotation invariant features are estimated from the noisy projections. These features are later on used to reconstruct the unknown image~\cite{zehni2019, zehni2020, Levin2018, Lingda2019}. Note that these methods require only one pass through the projection dataset and are therefore computationally more efficient. However, these methods are mainly used, when the underlying object is sparse~\cite{zehni2019, zehni2020}, projections in the form of tilt series are available~\cite{Lingda2019} or to recover a low-resolution ab-initio model~\cite{Levin2018}.  

\vspace{-20pt}
There is a recent surge in application of deep learning (DL) for the tomographic reconstruction task~\cite{Wang2020}. Majority of DL-based solutions assume the projection angles are known. These methods depending on the input to the DL model and the infusion of the physics of the problem can be broadly classified in three. In the first class, the network serves as an inverse operator that learns from large pools of supervised data to map projections (sinogram) back to image domain~\cite{Bo2018,Ge2020an}. These methods completely neglect any geometric knowledge of the problem and solely rely on the network and rich datasets to learn the underlying physics, which can be challenging for many inverse problems. In the second category of DL-based reconstruction methods, the geometry of the problem is taken into account~\cite{wurfl2016}. 
A plethora of recent methods deploy DL models to denoise initial FBP reconstructed images~\cite{jin2017, Quan2018,chen2017LDCT,Shan2019, han2018framelet,kang2018dlframe,yang2018WGAN, Zhong2020}.
DL-based completion or denoising of the sinograms in a low-dose computed tomography (CT) setting is also proposed in~\cite{dong2019, li2019inpaint}. The third category combines solving the optimization formulation of tomographic reconstruction along the gradient descent updates with machine learning components~\cite{Adler2017, Adler2018,Yang2020}. Also, deep image priors~\cite{Ulyanov_2018_CVPR} which benefit from the implicit prior offered by deep network architectures are recently adopted for tomographic reconstruction tasks~\cite{Baguer2020, Gong2019_dip}. 
While in the second and third categories the tomographic forward model is taken into account, they still heavily rely on the knowledge of the projection angles which might not be always available. However, here we address a regime where neither these projection angles nor their underlying probability distribution are known in advance.

\subsection{Contributions}
In this paper, we present an unsupervised adversarial learning based approach for tomographic reconstruction with unknown projection angles, namely \textit{UVTomo-GAN}. 
Unlike previous DL-based methods targeting the tomographic reconstruction problem, we address a more challenging inverse problem where the projection angles are unknown. Furthermore, our approach does not require large paired training sets and reconstructs an image given merely its tomographic measurements. By employing generative adversarial networks (GAN)~\cite{NIPS2014_5423}, our approach recovers the image and projection angle distribution through matching the distributions of the generated projections with the measurements.
Our proposed method is inspired by CryoGAN~\cite{cryogan} in which a 3D cryo-EM map is reconstructed given a large set of noisy projections. As opposed to CryoGAN, we consider a more challenging and realistic setting in which the projection angle distribution is not known a-priori, analogous to a cryo-EM set-up. Hence, we find the projection distribution alongside the 2D image. This work is an extension of our earlier paper~\cite{zehni2021uv}.

To recover the projection angle distribution in a GAN framework, we argue that the original generator's loss involves sampling from the projection angles distribution which is non-differentiable. To enable the flow of gradients in the backward pass through this non-differentiable operator, we alter the training loss at the generator side using Gumbel-Softmax approximation of samples from a categorical distribution~\cite{gumbelsoftmax}. Our proposed idea is general and applicable to a vast range of similar inverse problems such as~\cite{zehni2021}.

Furthermore, we adopt Hartley domain representation of the image expanded on a Hartley-Bessel (HB) basis in our reconstruction pipeline. 
Not only this truncated expansion represents a large class of images accurately, it also allows for the direct use of central slice theorem (CST) to generate the projections efficiently. %
Our theoretical analysis and simulation results affirm the ability of our method in recovering the image and projection distribution accurately from both clean and noisy measurements.  

The organization of this paper is as follows. We introduce the projection formation model and the reconstruction method in sections~\ref{sec:model} and~\ref{sec:method}. The analysis and experimental results are presented in~\ref{sec:analysis} and~\ref{sec:results}. We conclude the paper in~\ref{sec:conclusion}.

\vspace{-0.2cm}
\section{Projection Formation Model}
\label{sec:model}
We define the 1D projection formation model as,
\begin{equation}
    \zeta_{\ell} = \mathcal{P}_{\theta_{\ell}} I + \varepsilon_\ell, \quad \ell \in \{1, 2, ..., L\}
    \label{eq:proj_noisy}
\end{equation}
where $I: \mathbb{B}_2 \rightarrow \mathbb{R}_1$ is an unknown 2D compactly supported image in the unit ball $\mathbb{B}_2$ we wish to estimate. We restrict $I$ to the space of absolute and square integrable functions on $\mathbb{B}_2$, i.e., $I \in \mathcal{L}_1(\mathbb{B}_2) \cap \mathcal{L}_2(\mathbb{B}_2)$. $\mathcal{P}_{\theta}$ denotes the tomographic projection operator that takes the line integral along the parallel beams whose normal direction makes an angle $\theta  \in [0, 2\pi)$ with the $x$-axis,  
\begin{equation}
    (\mathcal{P}_{\theta} I) (x) = \int\limits_{-\infty}^{\infty} I(R_{\theta}\, \mathbf{x}) dy \label{eq:xray}
\end{equation}
where $\mathbf{x} = [x, y]^T$ represents the 2D Cartesian coordinates. 
$R_\theta$ is a $2 \times 2$ rotation matrix associated with $\theta$. As $I$ is compactly supported in $\mathbb{B}_2$, its projection along any direction would also be compactly supported in the unit ball, i.e., $\mathcal{P}_\theta I \in \mathcal{L}_1(\mathbb{B}_1) \cap \mathcal{L}_2(\mathbb{B}_1)$. We presume the projection angles $\{\theta_{\ell}\}_{\ell=1}^L$ are unknown and randomly drawn from an \textit{unknown} distribution $p$. Finally, the discretized projection lines of length $m$ are corrupted by additive white Gaussian noise $\varepsilon_\ell$ with zero mean and variance $\sigma^2$. Here we consider $\sigma$ to be known, although an unbiased estimator of $\sigma$ is attainable from the variance of the projections.

In this paper, given a large set of noisy projections, i.e., $\{\zeta_{\ell}\}_{\ell=1}^{L}$, we aim to recover the image $I$ and the unknown distribution of the projection angles $p$.

\vspace{-0.2cm}
\section{Method}
\label{sec:method}
\subsection{Image Representation}
To alleviate the computational cost of generating projections in practice,~\eqref{eq:proj_noisy} is evaluated in Fourier domain  using non-uniform fast Fourier transform~\cite{greengard2004accelerating} according to central slice theorem (CST). CST states that the Fourier transform of the projection corresponds to the central slice in the 2D Fourier domain,
\begin{equation}
\label{eq:central_slice}
\mathcal{F}(\mathcal{P}_{\theta} I )(\xi)  = \mathcal{F}(I)(\xi, \theta).
\end{equation}
with $\mathcal{F}$ denoting the Fourier transform and $(\xi, \theta)$ the polar coordinates. This motivates us to directly adopt CST to generate the projections. Therefore, in our pipeline we seek to recover the image in Fourier domain rather than pixel domain.

We use the Hartley transform of the images, which is a real representation closely related to Fourier transform and defined as:
\begin{align}
    \mathcal{H}(I) = \textrm{real}\{\mathcal{F}(I)\}-\textrm{imag}\{\mathcal{F}(I)\},
\end{align}
where $\mathcal{H}$ denotes the Hartley transform.
We assume the image $I$ has essential bandlimit $0 \! \leq \! s\leq \frac{1}{2}$ and is concentrated in the spatial domain with radius $R \leq \frac{m}{2}$. Therefore, $\mathcal{H}(I)$ can be expanded on an orthonormal basis on a disk of radius $s$. Based on the Fourier-Bessel basis introduced in~\cite{Zhao2016, Zhao:13}, we construct the real-valued steerable Hartley-Bessel (HB) basis $u_s^{k, q}(\xi,\theta) = J_s^{k, q}(\xi) \mathrm{cas}(k \theta)$ with radial functions
\begin{equation}
    \label{eq:radial_basis}
    J_s^{k, q}(\xi) = \begin{cases}
    N_{k, q} J_{k}\left(R_{k, q} \frac{\xi}{s}\right), & \xi \leq s, \\
    0, & \xi > s,
    \end{cases}
\end{equation}
where $J_{k}$ is the Bessel function of the first kind and integer order $k$, $R_{k, q}$ denotes the $q$-th root of $J_k$, and $N_{k,q}=(s \sqrt{\pi} \vert J_{k+1}(R_{k,q})\vert)^{-1}$ is the normalization factor. The angular part of the HB basis is $\textrm{cas}(k \theta) = \cos(k \theta) + \sin(k \theta)$.
We can expand $\mathcal{H}(I)$ on the HB basis, 
\begin{align}
    \mathcal{H}(I)(\xi, \theta) & =  \sum_{k=-\infty}^{\infty} \sum\limits_{q=1}^{\infty} c_{k, q} J_s^{k, q}\left(\xi\right) \textrm{cas}(k \theta). \label{eq:HB_expansion} 
\end{align}
Note that, $q$ and $k$ correspond to radial and angular frequencies. We can truncate the expansion in~\eqref{eq:HB_expansion} for functions that are well concentrated in real and Fourier space using a sampling criterion $R_{k, q} \leq 2\pi s R$~\cite{klug1972three,Zhao2016}. The maximum angular frequency index is denoted by $K_{\textrm{max}}$ and the maximum radial frequency for $k$-th angular frequency is denoted by $p_k$. The expansion coefficients $c= \{c_{k,q} \, \vert \, \forall (k, q) \; \textrm{s.t.} \; \vert k\vert \leq K_{\textrm{max}}, 1 \leq q \leq p_k\}$ are the unknown parameters of $I$ we aim to recover. For an image with $s\!<\!0.5$ or $R \!< \! \frac{m}{2}$, $c$ has less number of terms than the number of pixels $I$, i.e., the cardinality of $c < m^2$. Thus, $c$ would constitute a compressed representation of the image. 

Given the image expanded on HB basis, following CST, the Hartley transform of the projection from angle $\theta_\ell$ is simply obtained by setting $\theta=\theta_\ell$ in~\eqref{eq:HB_expansion} and is written as:
\begin{align}
    \mathcal{H}(\mathcal{P}_{\theta_\ell}\! I)\! (\xi)\! \!=\!  \sum_{k=-K_{\textrm{max}}}^{K_{\textrm{max}}} \! \sum\limits_{q=1}^{p_k} c_{k, q} J_s^{k,q}\!\left(\xi\right)\! \textrm{cas}(k \theta_\ell) \!=\! {H}_{\theta_\ell} (\xi) c.
    \label{eq:hartley_projection2} 
\end{align}
Therefore, we rewrite~\eqref{eq:proj_noisy} in Hartley domain as:
\begin{equation}
    \widetilde{{\zeta}}_{\ell} = {H}_{\theta_\ell} {c} + \widetilde{\varepsilon}_\ell, \, {\theta}_\ell \sim p, \quad {\ell} \in \{1, 2, ..., L\}, \label{eq:hartley_projs}
\end{equation}
\noindent with $\widetilde{\zeta} = \mathcal{H}(\zeta)$ and $\widetilde{\varepsilon} = \mathcal{H}(\varepsilon)$. The Hartley transform is unitary due to its self-adjoint and self-inverse properties. Therefore, the distribution of the Gaussian additive noise is preserved after taking the Hartley transform, i.e., $\widetilde{\varepsilon}_\ell \sim \mathcal{N}(\mathbf{0}_m, \sigma^2 I_m)$ where $\mathbf{0}_m$ is a vector of zeros of length $m$ and $I_m$ is an $m \times m$ identity matrix.

From the HB expansion coefficient $c$, we can reconstruct the image in the spatial domain,
\begin{align}
    I(r, \varphi) = \sum_{k=-K_{\textrm{max}}}^{K_{\textrm{max}}} \sum\limits_{q=1}^{p_k} c_{k, q} \, \mathcal{H} \left(u_s^{k,q}\right)(r, \varphi)
    \label{eq:HB_spatial}
\end{align}
where 
\begin{align}
    \mathcal{H}\!\left(u_s^{k,q}\right)\!(r, \varphi) \!=\! \frac{2 \sqrt{2 \pi} s (-1)^{(q + l)} R_{k,q} J_k(2 \pi s r)} {(2 \pi s r)^2 - R_{k, q}^2} \cos({k \varphi + \frac{\pi}{4}}),  \label{eq:inverse_basis}
\end{align}
and $l = \frac{k+1}{2}$ for odd $k$ and $l = \frac{k}{2}$ for even $k$. Since we have the analytical form of the basis function, we can easily evaluate the function values on Cartesian coordinates $[x, y]$ with $x=r \cos \varphi$ and $y = r \sin \varphi$.

\vspace{-0.2cm}
\subsection{Adversarial learning for reconstruction}
\label{sec:method_adv}
 \begin{figure}
     \centering
     \includegraphics[width=1 \linewidth]{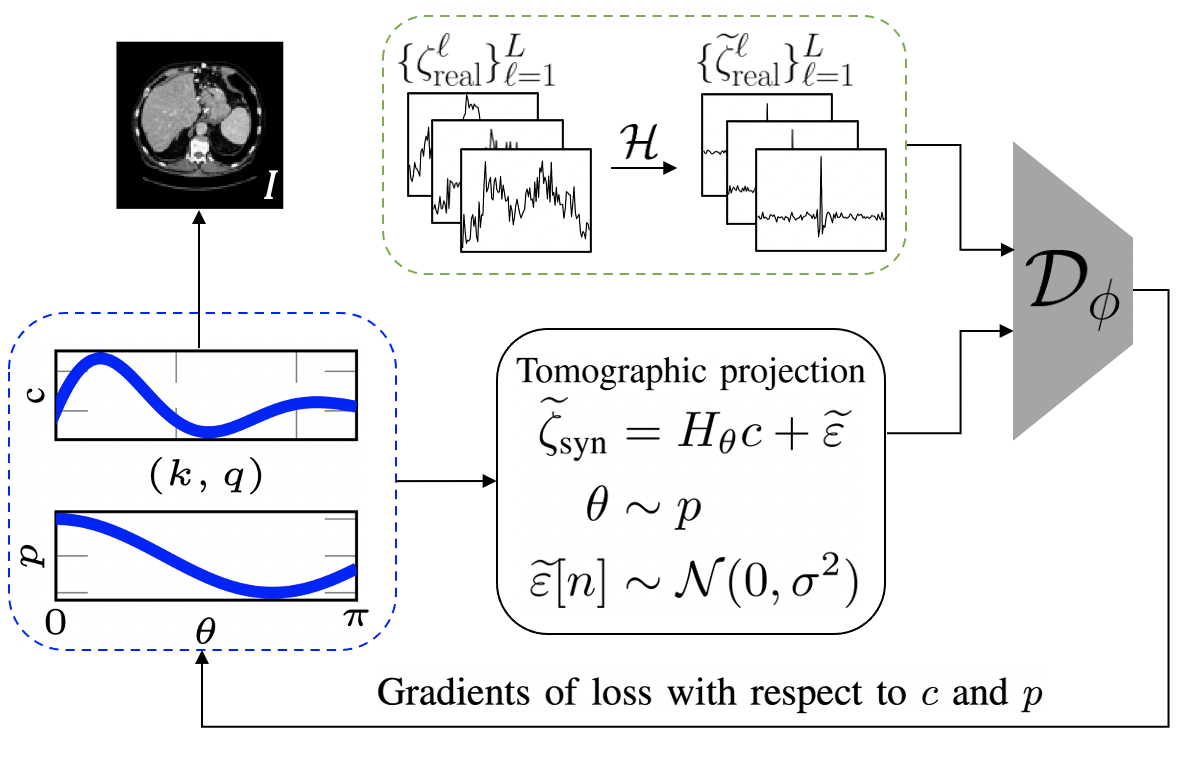}
     \vspace{-15pt}
     \caption{An illustration of our pipeline for adversarial learning based unknown view tomography reconstruction: Given the projections $\{\zeta^\ell_{\textrm{real}}\}_{\ell=1}^L$ (green dashed box), we recover the truncated Hartley-Bessel expansion coefficients $c$ of the image and projection angle distribution $p$ (blue dashed box).}
     \vspace{-15pt}
     \label{fig:pipeline}
 \end{figure}
 
Our reconstruction criterion is matching the distribution of the real projection dataset and the projections generated by $c$ and $p$ following~\eqref{eq:hartley_projs}. As GANs have proven suitable for matching a target distribution, we employ an adversarial learning framework presented in Fig.~\ref{fig:pipeline}.

Our adversarial learning approach consists of a critic $\mathcal{D}_{\phi}$ and a generator $\mathcal{G}$. Unlike classic GAN models with generators parameterized by neural networks with learnable weights, we specify the generator $\mathcal{G}$ by the known projection model defined in~\eqref{eq:hartley_projs}, the parameters of the image and projection angle distribution, i.e., $c$ and $p$. The generator's goal is to output projections that are close to the real projection dataset $\left \{\widetilde{\zeta}_{\textrm{real}}^{\ell} \right \}_{\ell=1}^L$ in distribution and hence fool the critic. For our model, the unknowns we seek to estimate at the generator side are $c$ and $p$. On the other hand, the critic $\mathcal{D}_{\phi}$, parameterized by $\phi$, tries to distinguish between the observations and the generated projections. Our pipeline is depicted in Fig.~\ref{fig:pipeline}.

We use Wasserstein GAN~\cite{wgan} loss with gradient penalty term (WGAN-GP)~\cite{wgangp}. We express the loss function in terms of $c$, $p$ and $\phi$ and the min-max problem as,
{\small
\begin{align}
    &\mathcal{L}(c, p, \phi) \! = \! \sum\limits_{b=1}^{B} \! \mathcal{D}_{\phi} \left(\widetilde{\zeta}^b_{\textrm{real}} \right) \! - \! \mathcal{D}_{\phi} \left (\widetilde{\zeta}^b_{\textrm{syn}}\right ) \!  - \! \lambda \! \left( \left \Vert \nabla_{\widetilde{\zeta}} \mathcal{D}_{\phi}\left(\widetilde{\zeta}^b_{\textrm{int}}\right) \right \Vert \! - \! 1 \! \right)^2  \label{eq:loss_function}\\
    &\widehat{c}, \widehat{p} = \argmin_{c, p} \max_{\phi} \mathcal{L}(c, p, \phi), \label{eq:minmax}
\end{align}}
where $\mathcal{L}$ denotes the loss, $B$ and $b$ represent the batch size and the index of a sample in the mini-batch, respectively. Also, $\widetilde{\zeta}_{\textrm{real}}$ and $\widetilde{\zeta}_{\textrm{syn}}$ mark the real and synthesized projections in Hartley domain. $\widetilde{\zeta}_{\textrm{syn}}$ is generated from the estimated image $\widehat{c}$ and projection distribution $\widehat{p}$ following $\widetilde{\zeta}_{\textrm{syn}} = {H}_{\theta} \widehat{c} + \widetilde{\varepsilon}$, $\theta \sim \widehat{p}$. Note that the last term in~\eqref{eq:loss_function} is the gradient penalty with weight $\lambda$ and roots from the Liptschitz continuity constraint of the critic in a WGAN setup. We use $\widetilde{\zeta}_{\textrm{int}}$ to denote a linearly interpolated sample between a real and a synthetic projection, i.e., $\widetilde{\zeta}_{\textrm{int}} = \alpha \, \widetilde{\zeta}_{\textrm{real}} + (1-\alpha) \, \widetilde{\zeta}_{\textrm{sim}}, \, \alpha \sim \textrm{Unif}(0, 1)$. In our experiments, we also used spectral normalization (SN)~\cite{miyato2018} and found that SN is a sufficient replacement for the gradient penalty term in terms of stabilizing the training. Thus, we set $\lambda=0$ in~\eqref{eq:loss_function} and only use spectral normalization to regularize the critic.
Following common practice, we solve~\eqref{eq:minmax} by alternating updates between $\phi$ and the generator's variables, i.e., $c$ and $p$, based on the associated gradients.

The loss at the generator side for a fixed $\mathcal{D}_{\phi}$ is,
\begin{equation}
    \mathcal{L}_{G}(c, p) = - \sum\limits_{b=1}^{B} \mathcal{D}_{\phi} ({H}_{\theta_b} c + \widetilde{\varepsilon}_b), \, \theta_b \sim p. 
    \label{eq:gen_loss}
\end{equation}
While~\eqref{eq:gen_loss} is differentiable with respect to $c$, its gradient of $p$ is not defined, as it involves sampling $\theta_b$ from the distribution $p$. This hinders updating $p$ through gradient back-propagation. To address this, we aim to design an alternative approximation of~\eqref{eq:gen_loss} which is differentiable with respect to $p$.

  \begin{algorithm}[t]
   \caption{UVTomo-GAN}\label{alg:ctgan}
     \textbf{Require:} $\alpha_\phi$, $\alpha_c$, $\alpha_p$: learning rates for $\phi$, $c$ and $p$. $n_{\textrm{disc}}$: the number of updates of the critic per generator update.\\ %
     \textbf{Input: } $ \left \{\widetilde{\zeta}^{\textrm{real}}_\ell \right \}_{\ell=1}^L$. Random initialization of $c$. The distribution $p$ is initialized with $\textrm{Unif}(0, 2\pi)$.\\
     \textbf{Output:} Estimates of $I$ and $p$. 
   \begin{algorithmic}[1]
     \While{$\phi$ has not converged}
     \For{$t=0,...,n_{\textrm{disc}}-1$}
     \State Sample a batch from real data, $\left \{\widetilde{\zeta}^b_{\textrm{real}} \right \}_{b=1}^B$
     \State Sample a batch of simulated projections using estimated $c$ and $p$, i.e. $ \left \{\widetilde{\zeta}^b_{\textrm{syn}} \right \}_{b=1}^B$ following~\eqref{eq:hartley_projs}
     \State Generate interpolated samples $ \left \{\widetilde{\zeta}^b_{\textrm{int}} \right \}_{b=1}^B$, $\widetilde{\zeta}^b_{\textrm{int}} = \alpha \, \widetilde{\zeta}^b_{\textrm{real}} + (1-\alpha) \, \widetilde{\zeta}^b_{\textrm{syn}}$ with $\alpha \sim \textrm{Unif}(0, 1)$
     \State Update the \textcolor{red}{critic} using gradient ascent steps using the gradient of~\eqref{eq:loss_function} with respect to $\phi$.
    \EndFor
     \State Sample a batch of $\{r_{i, b}\}_{b=1}^B$  using~\eqref{eq:gumbel_approx}
     \State Update $c$ and $p$ using stochastic gradient descent steps by taking the gradients of~\eqref{eq:total_loss} with respect to $c$ and $p$.
     \EndWhile
   \end{algorithmic}
 \end{algorithm}

To accommodate this approximation, we first discretize the support of the projection angles, i.e., $[0, 2\pi)$ into $N_{\theta}$ equal-sized bins. This makes $p$ a probability mass function (PMF) of length $N_\theta$ with the following properties:
\begin{align}
    \sum\limits_{i=0}^{N_\theta-1} p_i = 1, \textrm{and } p_i \geq 0, \forall i\in \{0, ..., N_\theta-1\}.
\end{align}
Now $p$ corresponds to a discrete or categorical distribution over $\theta$, which implies the sampled projection angles from $p$ can only belong to $N_{\theta}$ discrete categories. Therefore, we re-write the loss function~\eqref{eq:gen_loss} as:
\begin{equation}
    \mathcal{L}_{G}(c, p) = - \sum\limits_{b=1}^{B} \sum\limits_{t=0}^{N_\theta-1} \delta(\theta_t-\theta_b) \mathcal{D}_{\phi} ({H}_{\theta_t} c + \widetilde{\varepsilon}_b), \, \theta_b \sim p. 
    \label{eq:gen_loss_delta}
\end{equation}
A closer look at~\eqref{eq:gen_loss_delta} reveals that $\delta(\theta_t-\theta_b)$, $\theta_b \sim p$ is a sample from the discrete distribution $p$.
This enables us to incorporate the notion of Gumbel-Softmax  distribution and approximate~\eqref{eq:gen_loss} as:
\begin{equation}
    \mathcal{L}_{G}(c, p) \approx - \sum\limits_{b=1}^{B} \sum\limits_{i=0}^{N_\theta-1} r_{i, b}(p) \mathcal{D}_{\phi} ({H}_{\theta_i} c + \widetilde{\varepsilon}_b),
    \label{eq:gen_aprox_loss}
\end{equation}
with
\begin{align}
    r_{i, b}(p) \! = \! \frac{\exp{((g_{b,i} + \log(p_i))/\tau)}}{\sum\limits_{j=0}^{N_{\theta}-1} \exp{((g_{b,j} \! + \! \log(p_j))/\tau)}}, \, g_{b,i} \! \sim \! \textrm{Gumbel}(0, 1),
    \label{eq:gumbel_approx}
\end{align}
where $\tau$ is the softmax temperature factor. As $\tau \rightarrow 0$, $r_{i, b}(p)\! \rightarrow \! \textrm{one-hot} \left(\argmax_i [g_{b,i} \! + \! \log(p_i)] \right)$. Moreover, to obtain samples from the $\textrm{Gumbel}(0, 1)$ distribution, it suffices to draw $u \sim \textrm{Unif}(0, 1)$, $g \! = \! -\log(-\log(u))$~\cite{gumbelsoftmax}. Note that due to the reparametrization trick applied  in~\eqref{eq:gen_aprox_loss}, the approximated generator's loss has a tangible gradient with respect to $p$.

We also add prior knowledge on the image and projection distribution in the form of regularization terms. Hence, the regularized loss function we optimize at the generator side is:
\begin{align}
\mathcal{L}(c, p) \! &= \! \mathcal{L}_G(c, p) \! + \! \gamma_1 g_{\textrm{TV}}(c) \! + \! \gamma_2 \Vert c \Vert^2 \! + \! \gamma_3 \textrm{TV}(p) \! + \! \gamma_4 \Vert p \Vert^2
\label{eq:total_loss}
\end{align}

\noindent where we include total variation ($\textrm{TV}$) and $\ell_2$ regularization terms for the image, with $\gamma_1$ and $\gamma_2$ weights. To construct the TV of the image in terms of $c$, we use~\eqref{eq:HB_spatial} to render $I$ on a Cartesian grid in spatial domain and then compute total variation of $I$. Furthermore, we assume that the unknown PMF is a piece-wise smooth function of projection angles (which is a valid assumption especially in single particle analysis in cryo-EM~\cite{punjani2017}), therefore adding $\textrm{TV}$ and  $\ell_2$ regularization terms for the PMF with $\gamma_3$ and $\gamma_4$ weights. 
We present the pseudo-code for UVTomo-GAN in Alg.~\ref{alg:ctgan}. 

\vspace{-0.2cm}
\subsection{Maximum Marginalized Likelihood Estimation via Expectation-Maximization}
As a baseline for UVTomo-GAN, we consider maximum marginalized likelihood estimation (MMLE). We solve MMLE in Fourier domain via expectation-maximization (EM) and represent $\mathcal{F}(I)$ with its expansion coefficients $a$ on Fourier-Bessel bases. Thus, MMLE is formulated as 
\begin{align}
   \widehat{a}, \widehat{p} \!=\! \argmax_{a, p} \, \! \sum\limits_{\ell=1}^L \log \! \left(\sum\limits_{i=0}^{N_{\theta}-1} \! {P}(\mathcal{F}(\zeta_{\ell}) \vert a, \, \theta_i) p_i \right) \!. \label{eq:mmle_log}
\end{align}
To solve~\eqref{eq:mmle_log}, we take the gradients with respect to $a$ and $p$ and set them to zero. For $p$, we further impose $\sum_{i=0}^{N_\theta-1} p_i = 1$.
This yields the following alternating updates for $a$ and $p$, in the form of:
\begin{align}
    &\textrm{(E-step)}: r^t_{i,j} = \frac{\exp{\left(- \frac{\Vert \mathcal{F}(\zeta_i) - H_{\theta_j} a^{t-1} \Vert^2}{2 \sigma^2}\right)}}{\sum\limits_{j=0}^{N_\theta-1} p^{t-1}_j \exp\left(-\frac{\Vert \mathcal{F}(\zeta_i) - H_{\theta_j} a^{t-1} \Vert^2}{2 \sigma^2} \right) }, \label{eq:E-step}\\
    &(\textrm{M-step}): 
    \begin{cases}
    \bm{A}^t a^t = \bm{b}^t,   \\
    {p}_j^t \! = \! \frac{\sum\limits_{i=1}^{L} r_{i, j}^t}{\sum\limits_{i=1}^L \sum\limits_{j=0}^{N_{\theta}-1} r^t_{i, j}}, \label{eq:M-step}
    \end{cases}
\end{align}
where 
\begin{align}
        \bm{A}^t((k, q), (k', q')) &= \widehat{p^t}(k-k') \sum\limits_{\xi=1}^{N_\xi} J_{s}^{k, q}(\xi) J_{s}^{k', q'}(\xi) \\
        \widehat{p^t}(k) &= \sum\limits_{j=0}^{N_{\theta}-1} p_j^t \exp \left(- \imath \, \frac{2 \pi k j}{N_\theta} \right) \label{eq:A_def}
\end{align}
\begin{align}
     \bm{b^t}(k, q) &= \sum\limits_{\xi=1}^{N_\xi} \sum\limits_{j=0}^{N_{\theta}-1} J_{s}^{k, q}(\xi) \exp \left(- \imath \frac{2 \pi k j}{N_\theta} \right) \sum\limits_{i=1}^L r_{i, j} \mathcal{F}(\zeta_i)
\end{align}
where $r_{i, j}$ denotes the probability that the $i-$th projection is associated with $\theta_j$ angle and $t$ is the iteration index. Also, $H_\theta a$ generates the projection at $\theta$ direction in Fourier domain given FB expansion coefficients $a$. In~\eqref{eq:M-step}, $\bm{A}^t$ is indexed by $(k, q)$ and $(k', q')$ pairs and the discretization in $\xi$ is identical to the projection dataset. The advantages of using truncated FB expansion is that: (1) similar to HB representation, it provides an implicit regularization on the image, and (2) building matrix $\bm{A}^t$ in~\eqref{eq:E-step} in each iteration only requires rescaling the entries of a pre-computed matrix $\bm{J}((k, q), (k', q')) = \sum\limits_{\xi=1}^{N_\xi} J_{s}^{k, q}(\xi) J_{s}^{k', q'}(\xi)$ by $\widehat{p^t}(k -k')$. 

In~\eqref{eq:E-step}-\eqref{eq:M-step}, we update the probabilistic angular assignments for the projections in the E-step while updating $a$ and $p$ in the M-step. Note that, in the absence of noise, i.e., $\sigma=0$, the E-step reduces to template matching~\cite{Barnett2016}. To solve $a^t$ from the equation $\bm{A}^t a^t = \bm{b}^t$, we use preconditioned conjugate gradient descent~\cite{shewchuk1994}.

\subsection{Computational complexity}
We conclude this section by comparing the computational complexity per iteration of UVTomo-GAN and EM.

\noindent \textbf{UVTomo-GAN Complexity}: Based on Algorithm 1, we split the computational cost of UVTomo-GAN between: 1) the critic and 2) the generator (i.e., $c$ and $p$) updates. Let $C_\mathcal{D}$ denote a fixed computational cost related to forward and backpropagation passes through the critic $\mathcal{D}_\phi$. As expected, $C_{\mathcal{D}}$ depends on the batch size, network architecture and the size of its input. Thus, the larger the critic network, the higher the $C_{\mathcal{D}}$. For our critic architecture, we use a cascade of $N \ll m$ fully connected (FC) layers with intermediate ReLU non-linearities. Therefore, $C_{\mathcal{D}}$ points to the cost of matrix multiplication and backward passes through these $N$ layers. Furthermore, we keep the input and output sizes of these FC layers to be $O(m)$ ($m$ is the image/projection size). Therefore, $C_{\mathcal{D}} = O(m^2 N) = O(m^2)$. As these operations can be parallelized on GPU, forward and backward passes through $\mathcal{D}_\phi$ are time-efficient.
For batch size $B=O(m)$, the cost of critic update is $O(B \, C_{\mathcal{D}})=O(m^3)$. 

For updating the generator according to~\eqref{eq:gen_aprox_loss}, first we generate $N_{\theta} = O(m)$ projections or templates. This is done in $O(m^3)$. A thorough discussion on the derivation of this computational complexity term is deferred to Appendix~\ref{sec:cost_app}. %

In our implementation of~\eqref{eq:gen_aprox_loss}, instead of using $B$ different noise realizations $\{\widetilde{\varepsilon}_b\}_{b=1}^B$ for each of the clean templates, we consider $N_\theta$ noisy templates in total. This means the loss function we use at the generator side is:
\begin{align}
    \mathcal{L}_{G}(c, p) \approx - \sum\limits_{b=1}^{B} \sum\limits_{i=1}^{N_\theta} r_{i, b}(p) \mathcal{D}_{\phi} ({H}_{\theta_i} c + \widetilde{\varepsilon}_i). \label{eq:modified loss}
\end{align}
Indeed in the absence of noise, ~\eqref{eq:modified loss} matches~\eqref{eq:gen_aprox_loss}. However, in the noisy case, the benefits of~\eqref{eq:modified loss} are two-fold: 1) having the same performance as~\eqref{eq:gen_aprox_loss} empirically, 2) reducing the number of passes through the critic. 

Consequently, adding up the cost of passing $N_\theta$ projection templates through $\mathcal{D}_\phi$ leads to a total computational cost of $O(m^3 + m C_{\mathcal{D}})$ per generator update step. We update $c$ and $p$ every $n_{\textrm{disc}}$ iterations. Therefore, the average cost of UVTomo-GAN per iteration including the generator and critic's updates is $O(\frac{ (n_{\textrm{disc}}-1) m^3 + (m^3 + m C_{\mathcal{D}})}{n_{\textrm{disc}}}) = O(m^3)$.
\vspace{5pt}

\noindent \textbf{EM Complexity}: 
For EM, we specify the computational cost of E-step and M-step. At each E-step, we generate $N_\theta$ projection templates. If these templates are generated following CST and using the non-uniform Fourier transform of the image, they require $O(m^2 \log m)$ computations. Next, we update the angular assignments of $L$ projections by comparing them against $O(m)$ templates, hence a cost of $O(m^2 L)$. Then, the total cost of E-step is $O(m^2 \log m + m^2 L) = O(m^2 L)$. For the M-step, computing $\bm{b}^t$ from the projections costs $O(m^2 L)$ (or $O(m \log m L)$ if using FFT) while updating FB coefficients $a$ in~\eqref{eq:M-step} using conjugate gradient descent has $O(\sqrt{\kappa} \omega)$ computational cost~\cite{shewchuk1994} where $\omega$ is the number of non-zeros of $\bm{A}^t$ and $\kappa$ is its condition number. Note that $\omega = O(\eta\, m^3)$ depends on the number of non-zero elements in $\widehat{p^t}$, i.e. $\eta$. If all entries in $\widehat{p^t}$ are non-zero ($\eta = O(m)$), then the M-step's computational cost is $O(\sqrt{\kappa}\, m^4)$.
Finally, the overall computational complexity for EM is $O(\sqrt{\kappa}\, \eta \, m^3 +m^2L)$.

In terms of convergence, we empirically observe that UVTomo-GAN requires more training iterations.
We attribute this to the difference between the convergences of stochastic gradient descent used in UVTomo-GAN versus full batch processing in EM.
On the other hand, we show that while UVTomo-GAN is robust to the choice of initialization, EM is likely to get stuck in a bad local solution with random initialization. This observation is also reported in cryo-EM settings in~\cite{punjani2017, SCHERES2012519}.
\vspace{-5pt}

\section{Analysis}
\label{sec:analysis}
In this section, we first define our notations and then formally state the  reconstruction guarantees of UVTomo-GAN.
\vspace{-20pt}
\subsection{Notations}
We assume the image $f \in \mathcal{L}_1(\mathbb{B}_2) \cap \mathcal{L}_2(\mathbb{B}_2)$ has a bandlimit $0< s\leq 0.5$ and compactly supported in the unit ball $\mathbb{B}_2$. In addition, $f \in \textrm{span} \{ u^s_{k,q}\}_ \Omega$, $\Omega = \{(k, q) \, \vert \, \vert k\vert \leq K_{\textrm{max}}, 1 \leq q \leq p_k\}$ with $u^s_{k,q} = J_s^{k, q}(\xi) \textrm{cas}(k \theta)$. Thus, the Hartley transform of $f$ is expanded on a HB basis set. A measurement $\zeta$ associated with the projection angle $\theta \sim p$ is $\zeta = \mathcal{P}_{\theta} f + \varepsilon$ with $\varepsilon[n] \sim q_\varepsilon$ denoting additive IID noise. We assume $q_\varepsilon$ has full support in Fourier domain, i.e., $\{\mathcal{F} q_\varepsilon\}(\omega) \neq 0, \, \forall \omega$. 

Let $\mathrm{O}(2)$ denote the group of all possible rotations and reflections, i.e., $\Gamma^T \Gamma = I$ and $\textrm{det}(\Gamma) = \pm 1$, $\forall \Gamma \in \mathrm{O}(2)$. The action of the $\mathrm{O}(2)$ group on $f$ is defined as,
\begin{align}
    (\Gamma f)(\mathbf{x}) = f(\Gamma^{-1} \mathbf{x}), \, \forall \Gamma \in \mathrm{O}(2) \label{eq:group_action}
\end{align}
where $\mathbf{x}=[x, y]$ denotes the Cartesian coordinate. On the other hand, the action of $\Gamma$ on a probability distribution $p$ defined over $[0, \pi]$ manifests as a combination of flip or circular shift. The group $\mathrm{O}(2)$ partitions the space of $\textrm{span} \{u^s_{k,q}\}_\Omega$ into a set of equivalence classes where $[f] = \{\Gamma f, \, \forall \Gamma \in \mathrm{O}(2)\}$. Let $P^{\textrm{clean}}_{f, p}$ and $P^{\textrm{noisy}}_{f, p}$ denote the probability distributions induced by clean and noisy projections, i.e., $\zeta_{\textrm{clean}} = \mathcal{P}_\theta f$ and $\zeta_{\textrm{noisy}} = \mathcal{P}_\theta f + \varepsilon$ with $\theta \sim p$, respectively.
\vspace{-10pt}
\subsection{Theoretical results}
Here we elaborate upon the theoretical reconstruction guarantees of our proposed method.

\noindent \textbf{Theorem 1}: Consider $f, g \in \mathcal{L}_1(\mathbb{B}_2) \cap \mathcal{L}_2(\mathbb{B}_2)$ and the associated bounded probability distributions $p_f$ and $p_g$ on the projection angles distributed in $[0, 2\pi)$. Then,
\begin{align}
    P^{\textrm{clean}}_{f, p_f} = P^{\textrm{clean}}_{g, p_g} \Rightarrow [f] = [g], \, [p_f] = [p_g].
\end{align}
Furthermore, if $f = \Gamma g$, $\Gamma \in \mathrm{O}(2)$, then $p_f = \Gamma p_g$. 

\noindent The proof is provided in Appendix~\ref{sec:proof_thm1}. Intuitively, Theorem 1 states that if $f$ and $g$ have the same induced clean projection distribution, then the underlying objects and projection distributions are equivalent up to a rotation and reflection. We link the proof of this theorem to unique angular recovery in unknown view random tomography~\cite{Basu2000_1, Basu2000}.

\noindent \textbf{Theorem 2}: Assume $f\in \mathcal{L}_1(\mathbb{B}_2) \cap \mathcal{L}_2(\mathbb{B}_2)$ denoting the ground truth (GT) image and $p$ representing the bounded GT probability distribution over the projection angles $\theta\in [0, 2\pi)$. Let $\widehat{f}$ and $\widehat{p}$ stand for the recovered image and the bounded probability distribution after the convergence of UVTomo-GAN. Consider the asymptotic case as $L \rightarrow \infty$. Then, 
\begin{align}
    P^{\textrm{noisy}}_{f, p} = P^{\textrm{noisy}}_{\widehat{f}, \widehat{p}} \Rightarrow \widehat{f} = \Gamma f, \quad \widehat{p} = \Gamma p, 
\end{align}
for a unique $\Gamma \in \mathrm{O}(2)$.

\noindent The proof is available in Appendix~\ref{sec:proof_thm2}. This theorem validates that upon the convergence of UVTomo-GAN in the presence of noise and infinite number of noisy projections, the GT image and projection angle distribution is recovered up to a rotation-reflection transformation. We defer the study of sample complexity of UVTomo-GAN with finite size projection dataset to future work.
\section{Numerical Results}
\label{sec:results}
\pgfplotsset{every x tick label/.append style={font=\tiny, yshift=0.5ex}}
\pgfplotsset{every y tick label/.append style={font=\tiny, xshift=0.5ex}}
\begin{figure}%
\centering
	\begin{tikzpicture}
		\begin{groupplot}[group style={group size= 3 by 2,horizontal sep=0.4cm, vertical sep=0.3cm},     legend pos= north east,
					 legend style={legend cell align=left},
					 grid=both,                         %
    				 height=3cm,width=4.cm,
    				 xmin=23,xmax=123,
					 ymin=0,ymax=120, 
					 ylabel near ticks, xlabel near ticks] 
		\nextgroupplot[yticklabels={$0$, $2$, $4$, $6$},ytick={0, 20, 40, 60},xticklabels={,,},ymin=-20,ymax=70, x label style={at={(axis description cs:0.5,0)},anchor=north}, title={\footnotesize{Lung}},ylabel={\footnotesize{Spatial}},] 
			\addplot[darkred] table[x=x, y=y, col sep=comma]{figs/pdf_imgs/lung_proj_noise2.dat};
			 \addplot[darkblue] table[x=x, y=y, col sep=comma]{figs/pdf_imgs/lung_proj_clean2.dat};
			 
		\nextgroupplot[xticklabels={,,},yticklabels={,,},ymin=-20,ymax=70, title={\footnotesize{Abdomen}}, x label style={at={(axis description cs:0.5,0)},anchor=north}] 	
			\addplot[darkred] table[x=x, y=y, col sep=comma]{figs/pdf_imgs/chest_proj_noise2.dat};
			\addplot[darkblue] table[x=x, y=y, col sep=comma]{figs/pdf_imgs/chest_proj_clean2.dat};
			
		\nextgroupplot[xticklabels={,,},yticklabels={,,},ymin=-20,ymax=70,x label style={at={(axis description cs:0.5,0)},anchor=north}, title={\footnotesize{Ribosome}}]
			\addplot[darkred] table[x=x, y=y, col sep=comma]{figs/pdf_imgs/rib_proj_noise2.dat};
 			\addplot[darkblue] table[x=x, y=y, col sep=comma]{figs/pdf_imgs/rib_proj_clean2.dat};
 			
		\nextgroupplot[yticklabels={$0$, $1$, $2$},ytick={0, 1000, 2000},ymin=-700,ymax=2300, xticklabels={$0$, $25$, $50$, $75$, $100$},xtick={23, 48, 73, 98, 123}, x label style={at={(axis description cs:0.5,-0.4)},anchor=north}, xlabel style={yshift=2ex}, ,ylabel={\footnotesize{Hartley}},ytick scale label code/.code={$10^3$}] 	
			\addplot[darkred] table[x=x, y=y, col sep=comma]{figs/pdf_imgs/lung_proj_noise_h2.dat};
			\addplot[darkblue] table[x=x, y=y, col sep=comma]{figs/pdf_imgs/lung_proj_clean_h2.dat};
			
		\nextgroupplot[xticklabels={$0$, $25$, $50$, $75$, $100$},xtick={23, 48, 73, 98, 123}, yticklabels={,,},ymin=-700,ymax=2300, x label style={at={(axis description cs:0.5,-0.4)},anchor=north}, yticklabels={,,}, xlabel style={yshift=2ex}] 	
			\addplot[darkred] table[x=x, y=y, col sep=comma]{figs/pdf_imgs/chest_proj_noise_h2.dat};
			\addplot[darkblue] table[x=x, y=y, col sep=comma]{figs/pdf_imgs/chest_proj_clean_h2.dat};
			
		\nextgroupplot[yticklabels={$0$, $1$, $2$},ytick={0, 1000, 2000}, xticklabels={$0$, $25$, $50$, $75$, $100$},xtick={23, 48, 73, 98, 123}, ymin=-700,ymax=2300, yticklabels={,,}, x label style={at={(axis description cs:0.5,0)},anchor=north}] 	
			\addplot[darkred] table[x=x, y=y, col sep=comma]{figs/pdf_imgs/rib_proj_noise_h2.dat};
			\addplot[darkblue] table[x=x, y=y, col sep=comma]{figs/pdf_imgs/rib_proj_clean_h2.dat};
			
		\end{groupplot}
	\end{tikzpicture}
\caption{Samples of clean (blue) and noisy (red) projection lines in spatial (first row) and Hartely (second row) domain. For noisy data $\text{SNR}=3$. 
} 
\vspace{-0.4cm}
\label{fig:projs}
\end{figure}
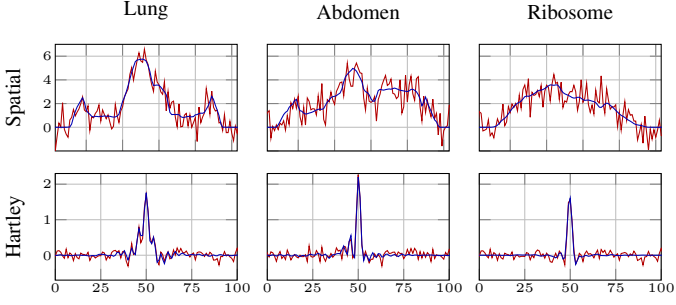

\input{figures/viz_imgs_clean_less_meas}

\input{figures/viz_imgs_snr3_final}
\pgfplotsset{every x tick label/.append style={font=\tiny, yshift=-0.1ex}}
\pgfplotsset{every y tick label/.append style={font=\tiny, xshift=-0.1ex}}
\begin{figure}[ht]%
\centering
	\begin{tikzpicture}
		\begin{groupplot}[group style={group size= 2 by 3,horizontal sep=0.2cm, vertical sep=0.6cm},     legend pos= north east,
		legend style={draw=none, fill=none, legend columns=-1, at={(1, 1.6)},anchor=north}, title style={at={(axis description cs:0.5, 0.85)}},
		grid=both,                         %
    	height=3cm,width=5.4cm,
    	xmin=1,xmax=101,
		ymin=0,ymax=1.2, 
		ylabel near ticks, xlabel near ticks] 
		
		\nextgroupplot[xticklabels={,,}, x label style={at={(axis description cs:0.5,0)},anchor=north}, title={\footnotesize{(a) Lung no noise}}, ylabel={\footnotesize{Intensity}}] 
			\addplot[darkblue,ultra thick] table[x=x, y=y, col sep=comma]{figs/line_profile/line_profile_lung_clean_gt.dat};
			\addplot[darkgreen,thick] table[x=x, y=y, col sep=comma]{figs/line_profile/line_profile_lung_clean_gl.dat};
            \addplot[darkred,thick] table[x=x, y=y, col sep=comma]{figs/line_profile/line_profile_lung_clean_made_gl.dat};
            \addplot[yellow,thick] table[x=x, y=y, col sep=comma]{figs/flips/line_profile_lung_clean_em_flip.dat};
            \addplot[black,thick] table[x=x, y=y, col sep=comma]{figs/flips/lung_clean_unknown_flip_line.dat};
            \legend{\footnotesize{GT}, \footnotesize{GLT}, \footnotesize{MADE+GL}, \footnotesize{EM}, \footnotesize{Ours}}

		\nextgroupplot[xticklabels={,,}, yticklabels={,,}, x label style={at={(axis description cs:0.5,0)},anchor=north}, title={\footnotesize{(b) Lung, $\textrm{SNR}=3$}}] 
			\addplot[darkblue,ultra thick] table[x=x, y=y, col sep=comma]{figs/line_profile/line_profile_lung_clean_gt.dat};
			\addplot[darkgreen,thick] table[x=x, y=y, col sep=comma]{figs/line_profile/line_profile_lung_mild_gl.dat};
            \addplot[yellow,thick] table[x=x, y=y, col sep=comma]{figs/flips/line_profile_lung_noise_em_flip.dat};
            \addplot[black,thick] table[x=x, y=y, col sep=comma]{figs/flips/lung_noise_unknown_flip_line.dat};
            
		\nextgroupplot[xticklabels={,,}, x label style={at={(axis description cs:0.5,0)},anchor=north}, title={\footnotesize{(c) Abdomen no noise}}, ylabel={\footnotesize{Intensity}}] 
			\addplot[darkblue,ultra thick] table[x=x, y=y, col sep=comma]{figs/line_profile/line_profile_chest_clean_gt.dat};
			\addplot[darkgreen,thick] table[x=x, y=y, col sep=comma]{figs/line_profile/line_profile_chest_clean_gl.dat};
            \addplot[darkred,thick] table[x=x, y=y, col sep=comma]{figs/line_profile/line_profile_chest_clean_gl_made.dat};
            \addplot[yellow,thick] table[x=x, y=y, col sep=comma]{figs/flips/line_profile_chest_clean_em_flip.dat};
            \addplot[black,thick] table[x=x, y=y, col sep=comma]{figs/flips/chest_clean_unknown_flip_line.dat};
            
		\nextgroupplot[xticklabels={,,}, yticklabels={,,}, x label style={at={(axis description cs:0.5,0)},anchor=north}, title={\footnotesize{(d) Abdomen, $\textrm{SNR}=3$}}] 
			\addplot[darkblue,ultra thick] table[x=x, y=y, col sep=comma]{figs/line_profile/line_profile_chest_clean_gt.dat};
			\addplot[darkgreen,thick] table[x=x, y=y, col sep=comma]{figs/line_profile/line_profile_chest_mild_gl.dat};
            \addplot[yellow,thick] table[x=x, y=y, col sep=comma]{figs/flips/line_profile_chest_noise_em_flip.dat};
            \addplot[black,thick] table[x=x, y=y, col sep=comma]{figs/flips/chest_noise_unknown_flip_line.dat};
            
		\nextgroupplot[x label style={at={(axis description cs:0.5,-0.1)},anchor=north}, title={\footnotesize{(e) Rib. no noise}}, xlabel={\footnotesize{Position}}, ylabel={\footnotesize{Intensity}}] 
			\addplot[darkblue,ultra thick] table[x=x, y=y, col sep=comma]{figs/line_profile/line_profile_rib_clean_gt.dat};
			\addplot[darkgreen,thick] table[x=x, y=y, col sep=comma]{figs/line_profile/line_profile_rib_clean_gl.dat};
            \addplot[darkred,thick] table[x=x, y=y, col sep=comma]{figs/line_profile/line_profile_rib_clean_gl_made.dat};
            \addplot[yellow,thick] table[x=x, y=y, col sep=comma]{figs/flips/line_profile_rib_clean_em_flip.dat};
            \addplot[black,thick] table[x=x, y=y, col sep=comma]{figs/flips/rib_clean_unknown_flip_line.dat};

		\nextgroupplot[yticklabels={,,}, x label style={at={(axis description cs:0.5,-0.1)},anchor=north}, title={\footnotesize{(f) Rib., $\textrm{SNR}=3$}}, xlabel={\footnotesize{Position}}] 
			\addplot[darkblue,ultra thick] table[x=x, y=y, col sep=comma]{figs/line_profile/line_profile_rib_clean_gt.dat};
			\addplot[darkgreen,thick] table[x=x, y=y, col sep=comma]{figs/line_profile/line_profile_rib_mild_gl.dat};
            \addplot[yellow,thick] table[x=x, y=y, col sep=comma]{figs/flips/line_profile_rib_noise_em_flip.dat};
            \addplot[black,thick] table[x=x, y=y, col sep=comma]{figs/flips/rib_noise_unknown_flip_line.dat};

		\end{groupplot}
	\end{tikzpicture}
\caption{Comparison between the line profile (middle vertical slice) of GT (blue) versus 1) GLT~\cite{Singer2013} (green), 2) MADE~\cite{Phan2017} + GL (red), 3) EM (yellow), 4) UVtomo-GAN with jointly optimizing $c$ and $p$ (black).
}
\label{fig:line_profiles}
\vspace{-0.5cm}
\end{figure}

\pgfplotsset{every x tick label/.append style={font=\tiny, yshift=-0.1ex}}
\pgfplotsset{every y tick label/.append style={font=\tiny, xshift=-0.1ex}}
\begin{figure}[ht]%
\centering
	\begin{tikzpicture}
		\begin{groupplot}[group style={group size= 2 by 3,horizontal sep=0.7cm, vertical sep=0.8cm},     legend pos= north east,
					 legend style={legend cell align=left},
					 legend style={at={(1, 1)}, legend cell align=left, draw=none, fill=none},
					 grid=both,                         %
    				 height=3.1cm,width=5.1cm,
    				 xmin=1,xmax=120,
					 ymin=0,ymax=0.003, 
					 ylabel near ticks, xlabel near ticks, ytick={0, 0.01, 0.02}, yticklabels={0, 0.5, 1},] 
		\nextgroupplot[xticklabels={,,},ylabel={PMF},ymin=0,ymax=0.02,  x label style={at={(axis description cs:0.5,0)},anchor=north}, 
		title={\footnotesize{(a) Lung-No noise, $d_\textrm{TV}\!=\!0.041$}},
		] 
			\addplot[darkblue,very thick] table[x=x, y=y, col sep=comma]{figs/pdf_imgs/pmf_lung_gt_less_meas.dat};
			\addplot[darkgreen,ultra thick] table[x=x, y=y, col sep=comma]{figs/pdf_imgs/pmf_lung_gt_clean_100k_less_meas.dat};
			\addplot[darkred,very thick] table[x=x, y=y, col sep=comma]{figs/flips/lung_clean_unknown_flip_pmf_est.dat};

		\nextgroupplot[yticklabels={,,},xticklabels={,,},ymin=0,ymax=0.02, x label style={at={(axis description cs:0.5,0)},anchor=north}, ytick={0, 0.01, 0.02},
		title={\footnotesize{(b) Lung-SNR $3$, $d_\textrm{TV}\!=\!0.073$}},
		ytick scale label code/.code={}] 
			\addplot[darkblue,very thick] table[x=x, y=y, col sep=comma]{figs/pdf_imgs/pmf_lung_noise_gt.dat};
			\addplot[darkgreen,ultra thick] table[x=x, y=y, col sep=comma]{figs/pdf_imgs/pmf_lung_gt_clean_100k_less_meas.dat};
            \addplot[darkred,very thick] table[x=x, y=y, col sep=comma]{figs/flips/lung_noise_unknown_flip_pmf_est.dat};
			\legend{\tiny{Sample $p$},\tiny{Origin. $p$}, \tiny{$\widehat{p}$}}

		\nextgroupplot[xticklabels={,,},ylabel={PMF},ymin=0,ymax=0.022, 
		title={\footnotesize{(c) Abdomen-No noise, $d_\textrm{TV}\!=\!0.040$}}, ytick={0, 0.01, 0.02}, yticklabels={0, 0.5, 1},
		xticklabels={,,}, x label style={at={(axis description cs:0.5,-0.4)},anchor=north}, xlabel style={yshift=2ex}] 	
			\addplot[darkblue,very thick] table[x=x, y=y, col sep=comma]{figs/pdf_imgs/pmf_chest_gt_less_meas.dat};
			\addplot[darkgreen,ultra thick] table[x=x, y=y, col sep=comma]{figs/pdf_imgs/pmf_chest_gt_origin.dat};
            \addplot[darkred,very thick] table[x=x, y=y, col sep=comma]{figs/flips/chest_clean_unknown_flip_pmf_est.dat};

		\nextgroupplot[xticklabels={,,},yticklabels={,,},ymin=0,ymax=0.02, x label style={at={(axis description cs:0.5,-0.4)},anchor=north},
		title={\footnotesize{(d) Abdomen-SNR $3$, $d_\textrm{TV}\!=\!0.080$}}, ytick={0, 0.01, 0.02}, xticklabels={,,}, yticklabels={,,}, ytick scale label code/.code={},] 	
			\addplot[darkblue,very thick] table[x=x, y=y, col sep=comma]{figs/pdf_imgs/pmf_chest_noise_gt.dat};
			\addplot[darkgreen,ultra thick] table[x=x, y=y, col sep=comma]{figs/pdf_imgs/pmf_chest_gt_origin.dat};
			\addplot[darkred,very thick] table[x=x, y=y, col sep=comma]{figs/flips/chest_noise_unknown_flip_pmf_est.dat};

		\nextgroupplot[ylabel={PMF},ymin=0,ymax=0.02,
		ytick={0, 0.01, 0.02}, yticklabels={0, 0.5, 1},
		title={\footnotesize{(e) Rib.-No noise, $d_\textrm{TV}\!=\!0.041$}},
		xtick={1,30, 60, 90, 120}, xticklabels={$0$,$\frac{\pi}{4}$,$\frac{\pi}{2}$,$\frac{3\pi}{4}$,$\pi$}, xlabel={\footnotesize{Projection angle $\theta$}}, xlabel style={yshift=1.5ex}] 	
			\addplot[darkblue,very thick] table[x=x, y=y, col sep=comma]{figs/pdf_imgs/pmf_rib_gt_less_meas.dat};
			\addplot[darkgreen,ultra thick] table[x=x, y=y, col sep=comma]{figs/pdf_imgs/pmf_rib_gt_origin.dat};
            \addplot[darkred,very thick] table[x=x, y=y, col sep=comma]{figs/flips/rib_clean_unknown_flip_pmf_est.dat};

		\nextgroupplot[yticklabels={,,},ymin=0,ymax=0.02, ytick={0, 0.01, 0.02}, 
		title={\footnotesize{(f) Rib.-SNR $3$, $d_\textrm{TV}\!=\!0.070$}},
		ytick scale label code/.code={},  xlabel={\footnotesize{Projection angle $\theta$}}, xtick={1,30, 60, 90, 120}, xticklabels={$0$,$\frac{\pi}{4}$,$\frac{\pi}{2}$,$\frac{3\pi}{4}$,$\pi$}, xlabel style={yshift=1.5ex}] 	
			\addplot[darkblue,very thick] table[x=x, y=y, col sep=comma]{figs/pdf_imgs/pmf_rib_gt.dat};
			\addplot[darkgreen,ultra thick] table[x=x, y=y, col sep=comma]{figs/pdf_imgs/pmf_rib_gt_origin.dat};
			\addplot[darkred,very thick] table[x=x, y=y, col sep=comma]{figs/flips/rib_noise_unknown_flip_pmf_est.dat};
			
		\end{groupplot}
	\end{tikzpicture}
\caption{Comparison between the original GT $p$ (green) used to sample the projection angles from, the empirical sample distribution of the projection angles (blue) and the one estimated by our method $\widehat{p}$ (red). Within each row, the subplots share the same vertical axis. All subplots have the same legends. For no noise settings, $d_{TV}$ is computed between $\widehat{p}$ (red) and original $p$ (green), while for the noisy case, $d_{TV}$ is computed between $\widehat{p}$ (red) and sample estimation of $p$ (blue).
}
\label{fig:pmf_results}
\vspace{-0.5cm}
\end{figure}

\pgfplotsset{every x tick label/.append style={font=\tiny, yshift=0.1ex}}
\pgfplotsset{every y tick label/.append style={font=\tiny, xshift=0.1ex}}
\begin{figure}%
\centering
	\begin{tikzpicture}
		\begin{groupplot}[group style={group size= 2 by 3,                      %
    				horizontal sep=0.25cm, vertical sep=0.2cm}, 
    				legend pos= south west,        %
					 legend style={at={(0.25,-0.1)}, legend cell align=left, draw=none, fill=none},
					 legend style={legend cell align=left},
					 grid=both,                         %
    				 height=3cm,width=5.3cm,
    				 xmin=0,xmax=20,
					 ymin=5,ymax=30, 
					 ylabel near ticks, xlabel near ticks] 
		\nextgroupplot[xmin=1, xmax=20, ymin=5, ymax=30, x label style={at={(axis description cs:0.5,-0.2)},anchor=north}, ylabel={\tiny{PSNR(dB)-Lung}}, ylabel style={yshift=-1ex}, title=\footnotesize{No noise}, ,title style={yshift=-1.2ex}, xticklabels={,,}, xmode=log] 	
		\addplot[darkblue,very thick] table[x=x, y=y, col sep=comma]{figs/flips/lung_clean_known_flip_psnr.dat};
		\addplot[darkred,very thick] table[x=x, y=y, col sep=comma]{figs/flips/lung_clean_fixed_flip_psnr.dat};
		\addplot[darkgreen,very thick] table[x=x, y=y, col sep=comma]{figs/flips/lung_clean_unknown_flip_psnr.dat};
		\legend{{\tiny Given $p$},{\tiny Assume $p$ Unif.}, {\tiny Jointly optimize $c$ and $p$}}
		
		\nextgroupplot[xmin=1, xmax=20, ymin=5, ymax=30, x label style={at={(axis description cs:0.5,-0.2)},anchor=north}, title=\footnotesize{$\text{SNR}=3$},title style={yshift=-1.2ex},xticklabels={,,}, yticklabels={,,}, xmode=log] 	
		\addplot[darkblue,very thick] table[x=x, y=y, col sep=comma]{figs/flips/lung_noise_known_flip_psnr.dat};
		\addplot[darkred,very thick] table[x=x, y=y, col sep=comma]{figs/flips/lung_noise_fixed_flip_psnr.dat};
		\addplot[darkgreen,very thick] table[x=x, y=y, col sep=comma]{figs/flips/lung_noise_unknown_flip_psnr.dat};
		
		\nextgroupplot[xmin=1, xmax=20, ymin=5, ymax=32, x label style={at={(axis description cs:0.5,-0.2)},anchor=north}, xticklabels={,,}, ylabel style={yshift=-1ex}, xlabel style={yshift=0.8ex}, ylabel={\tiny{PSNR(dB)-Abdomen}}, xmode=log]
		\addplot[darkblue,very thick] table[x=x, y=y, col sep=comma]{figs/flips/chest_clean_known_flip_psnr.dat};
		\addplot[darkred,very thick] table[x=x, y=y, col sep=comma]{figs/flips/chest_clean_fixed_flip_psnr.dat};
		\addplot[darkgreen,very thick] table[x=x, y=y, col sep=comma]{figs/flips/chest_clean_unknown_flip_psnr.dat};
		
		\nextgroupplot[xmin=1, xmax=20, ymin=5, ymax=32, x label style={at={(axis description cs:0.5,-0.2)},anchor=north}, yticklabels={,,}, xticklabels={,,}, xmode=log] 	
		\addplot[darkblue,very thick] table[x=x, y=y, col sep=comma]{figs/flips/chest_noise_known_flip_psnr.dat};
		\addplot[darkred,very thick] table[x=x, y=y, col sep=comma]{figs/flips/chest_noise_fixed_flip_psnr.dat};
		\addplot[darkgreen,very thick] table[x=x, y=y, col sep=comma]{figs/flips/chest_noise_unknown_flip_psnr.dat};
		
		\nextgroupplot[xmin=1, xmax=20, ymin=5, ymax=35, x label style={at={(axis description cs:0.5,-0.2)},anchor=north}, ylabel={\tiny{PSNR (dB)-Rib.}}, ylabel style={yshift=-1ex}, xlabel={\footnotesize{$\times 10\textrm{k}$ iteration}}, xlabel style={yshift=0.8ex}, xmode=log] 	
		\addplot[darkblue,very thick] table[x=x, y=y, col sep=comma]{figs/flips/rib_clean_known_flip_psnr.dat};
		\addplot[darkred,very thick] table[x=x, y=y, col sep=comma]{figs/flips/rib_clean_fixed_flip_psnr.dat};
		\addplot[darkgreen,very thick] table[x=x, y=y, col sep=comma]{figs/flips/rib_clean_unknown_flip_psnr.dat};
		
		\nextgroupplot[xmin=1, xmax=20, ymin=5, ymax=35, x label style={at={(axis description cs:0.5,-0.2)},anchor=north},  xlabel={\footnotesize{$\times 10\textrm{k}$ iteration}}, xlabel style={yshift=0.8ex}, yticklabels={,,}, xmode=log] 	
		\addplot[darkblue,very thick] table[x=x, y=y, col sep=comma]{figs/flips/rib_noise_known_flip_psnr.dat};
		\addplot[darkred,very thick] table[x=x, y=y, col sep=comma]{figs/flips/rib_noise_fixed_flip_psnr.dat};
		\addplot[darkgreen,very thick] table[x=x, y=y, col sep=comma]{figs/flips/rib_noise_unknown_flip_psnr.dat};
		
		\end{groupplot}
	\end{tikzpicture}
\caption{Convergence results for the no noise and noisy ($\textrm{SNR}=3$) experiments. The setting of the experiments are the same as the ones in Fig.~\ref{fig:viz_results_clean}-\ref{fig:viz_results}. We compare UVTomo-GAN jointly optimizing for $c$ and $p$ (green) versus GAN-based baselines with given $p$ (blue) and assumed $p$ as uniform distribution (red). Vertical axis shows the PSNR in dB and the horizontal axis is the training iteration number. The subplots in each row share the same vertical and horizontal axis.
}
\vspace{-0.4cm}
\label{fig:psnr_results}
\end{figure}

\input{figures/conv_imgs_less_meas}

\subsection{Experiment setup}
\noindent \textbf{Dataset:} In our experiments, to verify the generalization of our method, we use three different images. Two are biomedical images of lung and abdomen from low dose CT (LDCT) dataset~\cite{ldct}. For the third image, we generated the 3D map of 100S Ribosome~\cite{ribosome} using its protein sequence in Chimera~\cite{chimera} and took a 2D projection of the generated map along a random view. 
We resized all images to $101 \times 101$ dimension. We refer to these images as Lung, Abdomen and Ribosome. We synthesize the real projection dataset in Hartley domain  following~\eqref{eq:hartley_projs} where $p$ is a smooth probability distribution over the projection angles and is chosen randomly. To generate the real dataset, we finely discretize the projection angle domain $[0, \pi)$ with  $240$ equal sized bins and use non-uniform polar FFT~\cite{Averbuch2006} and CST to generate the projections. We also add the flipped projections to the dataset, such that $\theta$ covers $[0, 2\pi)$. This means $p$ has a symmetry where $p(\theta) = p(\theta+\pi)$, for $\theta \in [0, \pi)$. Therefore, when recovering $p$ we only recover $p$ in $[0, \pi)$ range. Throughout this draft, we visualize $p$ on $[0, \pi)$. For the reconstruction, we consider a coarser grid for the projection angles with $N_\theta = 240$ bins for the interval $[0, 2\pi)$. This way we are taking into account the approximated discretization of $\theta$ at the reconstruction time which might differ from how the real projection angles are obtained.   We study two noise regimes: 1) no noise, and 2) noisy with $\mathrm{SNR}=3$, $\mathrm{SNR}$ denoting the ratio of signal-to-noise variance of the projections:

\begin{align}
    \text{SNR} = \frac{\textrm{Var}\{ \zeta_{\textrm{clean}} \}}{\textrm{Var}\{\zeta_{\textrm{noisy}}-\zeta_{\textrm{clean}}\}}
\end{align}
where $\zeta_{\textrm{clean}}$ and $\zeta_{\textrm{noisy}}$ stand for the clean and noisy projections in spatial domain, respectively. Examples of clean and noisy projections in both spatial and Hartley domains are illustrated in Fig.~\ref{fig:projs}. In our experiments with clean data, the number of projections before adding the flipped versions is $L=2 \times 10^3$, while for noisy experiments, $L=2 \times 10^4$.

\noindent \textbf{Training and Network Architecture}: We set a batch-size of $B=200$. We fix the regularization weights on the PMF as $\gamma_3=0.01$ and $\gamma_4=0.04$ unless otherwise stated. For the lung and abdomen images in the clean case, the default image regularization weights are $\gamma_1=10^{-5}$ and $\gamma_2=5 \times 10^{-5}$ while having zero $\gamma_1$ and $\gamma_2$ for the Ribosome dataset. In the noisy case, to obtain the best results in various settings and take into account the difference in the projection datasets, we select $\gamma_1$ from $\{0.0005, 0.001, 0.002, 0.005 \}$ and $\gamma_2$ from $\{0.005, 0.02, 0.04\}$.

We have separate learning rates for $\mathcal{D}_{\phi}$, $c$ and $p$ denoted by $\alpha_{\phi}$, $\alpha_{c}$ and $\alpha_{p}$, but often choose $\alpha_{\phi}=\alpha_{{c}}$. We choose the initial values of $\alpha_{\phi}$, $\alpha_{c}$ and $\alpha_{p}$ from $[0.002, 0.01]$ with a step-decay schedule. 
We update $\mathcal{D}_\phi$, ${c}$ and $p$ using stochastic gradient descent (SGD) steps. We clip the gradients of $\mathcal{D}_\phi$ and ${c}$ by $1$ and $10$ respectively and normalize the gradients of $p$ to have norm $0.1$. We train the critic $n_{\textrm{disc}}\!=\!4$ times per updates of ${c}$ and $p$. Although, after training for a while, we increase the frequency of updating ${c}$ and $p$ by setting $n_{\textrm{disc}}=2$. Once converged, we use the reconstructed HB expansion coefficients to re-render the image in spatial domain according to~\eqref{eq:HB_spatial}.

Our architecture of the critic consists of four fully connected (FC) layers with $\ell$, $\ell/2$, $\ell/4$, and $1$ output sizes with ReLU~\cite{xu2015empirical} activations in between. We choose $\ell=512$ for no noise and $\ell=256$ for noisy experiments. Our justification for adopting a smaller critic network in noisy case is to avoid overfitting to noisy projections and reduce the leak of noise in the final reconstruction.

To improve the stability of the GAN training, we use spectral normalization~\cite{miyato2018}, applied to all critic layers. To enforce $p$ to have non-negative values while summing up to one, we set it to be the output of a $\texttt{Softmax}$ layer. To check the robustness of UVTomo-GAN with respect to initialization, in our experiments we try two initialization schemes for ${c}$: 1) initialize each entry of $c$ independently with a random variable drawn from $\mathcal{N}(0, 4 \times 10^{-4})$, 2) $c_{0, 0} = 0.01$ and $c_{k, q}=0, \, \forall \, k\neq0, q\neq0$. In this draft we only report the results of the first initialization scheme as both led to similar results. We set $p$ to be a uniform distribution initially. For the critic, we randomly initialize the weights of the FC layers with a zero-mean Gaussian distribution and standard deviation $0.05$ and set the biases to zero. Our implementation is in PyTorch and runs on single GPU.

\noindent \textbf{Evaluation metrics:} To assess the quality of the reconstructed image, we use peak signal to noise ratio (PSNR) and normalized cross correlation (CC). Higher value of these metrics signals better quality of the reconstruction. Also, to evaluate $\widehat{p}$ compared to the ground truth, we use total variation distance ($\textrm{TV}$) defined as:
\begin{align}
    d_\mathrm{TV} = \frac{1}{2} \Vert p - \widehat{p} \Vert_1.
\end{align}

\subsection{Baselines}
We benchmark UVTomo-GAN with unknown $p$ against five baselines. In our first baseline, namely graph Laplacian tomography (GLT), the projections with unknown views are sorted following~\cite{Singer2013} and the image is reconstructed accordingly. Note that compared to~\cite{Coifman2008}, \cite{Singer2013} is more resilient to noise. In our experiments, as we deal with non-uniform unknown projection angle distribution, sorting based methods struggle in reconstructing the image accurately (as shown in the Fig.~\ref{fig:viz_results_clean}-\ref{fig:viz_results}, second column), unless order statistics of the projection angle distribution is known~\cite{Coifman2008}. However, the order statistics are unknown as the projection angle distribution itself is unknown.

Thus, for our second baseline, we aim to recover the projection angles rather than their sorting order. As a result, we combine MADE~\cite{Phan2017} and spectral analysis of the graph Laplacian (GL) to obtain the angle corresponding to each projection. We name this baseline MADE+GL. Using the moment-based approach in~\cite{Phan2017}, we obtain the angular differences between any two projections. Next, after thresholding the angular differences, we construct a weight matrix similar to~\cite{Coifman2008}. Finally, we obtain the spectral embedding of the projections and consequently the projection angles after spectral decomposition of the normalized weight matrix. For both GLT and MADE+GL baselines, after the estimation of the projection ordering and angles, we reconstruct the image via a TV regularized optimization solved by ADMM~\cite{boyd2011distributed} using GlobalBioIm library~\cite{globalbioim}. 

As our third baseline, we compare against MMLE~\eqref{eq:mmle_log} solved by EM~\eqref{eq:E-step}-\eqref{eq:M-step}. We initialize EM with $10$ random initializations. We test two different forms of initializations, 1) randomly located Gaussian blobs with random standard deviations, 2) initializing each pixel with Uniform distribution within a circular mask, i.e., $I[x, y] \sim \textrm{Unif}[0, 1]$. In our experiments, we report the best results for EM out of these $10$ random initializations, hence the name \textit{EM best random init} for this baseline. Examples of initializations for EM are provided in Fig.~\ref{fig:em_init}.  We provide more details on our baselines in Appendix~\ref{sec:baseline_detail}
\pgfplotsset{colormap name=viridis}
\begin{figure}[H]
\begin{center}
\begin{tikzpicture}
\begin{groupplot}[group style={group size= 6 by 1, horizontal sep=0.05cm, vertical sep=0.1cm}, legend pos= north east]

\nextgroupplot[view={0}{90},
              xmin=0,
              xmax=63,
              ymin=0,
              ymax=63,
              height=3cm,width=3cm,
              legend style={at={(0.45,0.98)},draw=none, fill=none},
              xlabel near ticks,
	           ylabel near ticks,
              title style={yshift=-1.5ex,},
              xticklabels={,,},
              yticklabels={,,},
              enlargelimits=false, axis on top, axis equal image, ticks=none]
    \addplot graphics[xmin=0,ymin=0,xmax=63,ymax=63] {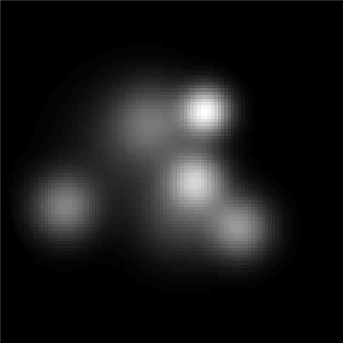};
               
    \nextgroupplot[view={0}{90},
              xmin=0,
              xmax=63,
              ymin=0,
              ymax=63,
              height=3cm,width=3cm,
              legend style={at={(0.45,0.98)},draw=none, fill=none},
              xlabel near ticks,
	           ylabel near ticks,
              title style={yshift=-1.5ex,},
              xticklabels={,,},
              yticklabels={,,},
              enlargelimits=false, axis on top, axis equal image, ticks=none]
    \addplot graphics[xmin=0,ymin=0,xmax=63,ymax=63] {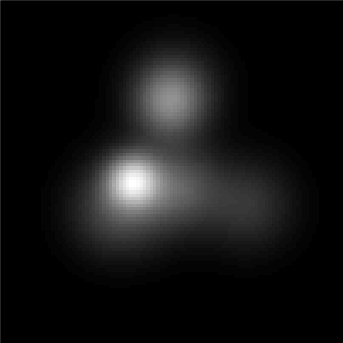};
    
    \nextgroupplot[view={0}{90},
              xmin=0,
              xmax=63,
              ymin=0,
              ymax=63,
              height=3cm,width=3cm,
              legend style={at={(0.45,0.98)},draw=none, fill=none},
              xlabel near ticks,
	           ylabel near ticks,
              title style={yshift=-1.5ex,},
              xticklabels={,,},
              yticklabels={,,},
              enlargelimits=false, axis on top, axis equal image, ticks=none]
    \addplot graphics[xmin=0,ymin=0,xmax=63,ymax=63] {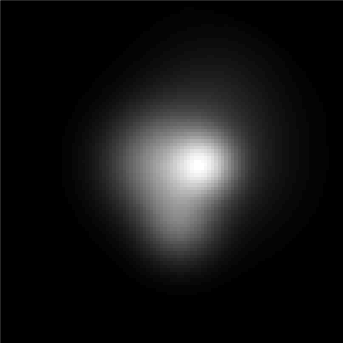};
    
    \nextgroupplot[view={0}{90},
              xmin=0,
              xmax=63,
              ymin=0,
              ymax=63,
              height=3cm,width=3cm,
              legend style={at={(0.45,0.98)},draw=none, fill=none},
              xlabel near ticks,
	           ylabel near ticks,
              title style={yshift=-1.5ex,},
              xticklabels={,,},
              yticklabels={,,},
              enlargelimits=false, axis on top, axis equal image, ticks=none]
    \addplot graphics[xmin=0,ymin=0,xmax=63,ymax=63] {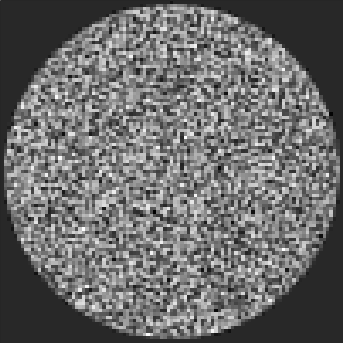};
               
    \nextgroupplot[view={0}{90},
              xmin=0,
              xmax=63,
              ymin=0,
              ymax=63,
              height=3cm,width=3cm,
              legend style={at={(0.45,0.98)},draw=none, fill=none},
              xlabel near ticks,
	           ylabel near ticks,
              title style={yshift=-1.5ex,},
              xticklabels={,,},
              yticklabels={,,},
              enlargelimits=false, axis on top, axis equal image, ticks=none]
    \addplot graphics[xmin=0,ymin=0,xmax=63,ymax=63] {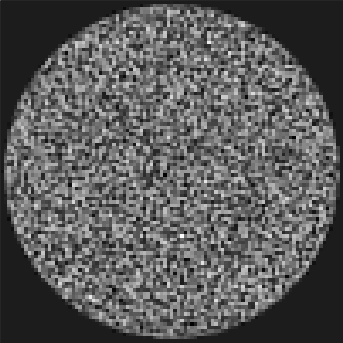};
    
    \nextgroupplot[view={0}{90},
              xmin=0,
              xmax=63,
              ymin=0,
              ymax=63,
              height=3cm,width=3cm,
              legend style={at={(0.45,0.98)},draw=none, fill=none},
              xlabel near ticks,
	           ylabel near ticks,
              title style={yshift=-1.5ex,},
              xticklabels={,,},
              yticklabels={,,},
              enlargelimits=false, axis on top, axis equal image, ticks=none]
    \addplot graphics[xmin=0,ymin=0,xmax=63,ymax=63] {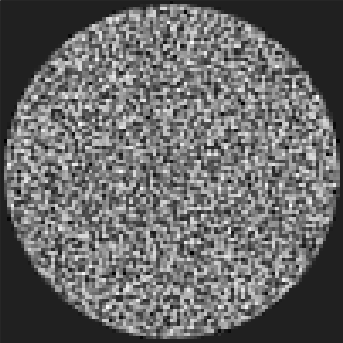};
  \end{groupplot}
\end{tikzpicture}
\caption{Examples of the initialization images used in EM. }
\label{fig:em_init}
\end{center}
\vspace{-0.5cm}
\end{figure}

To evaluate the effect of estimating the projection angle distribution $p$, we consider two GAN-based benchmarks. In the first, we assume that $p$ is given in advance. Note that, this baseline is the adaptation of CryoGAN~\cite{cryogan} (where the distribution of the latent variables is presumed to be known) to the 2D unknown view tomography problem. For our second GAN-based benchmark, we assume $p$ to be a uniform distribution. In both GAN-based baselines, we follow Alg.~\ref{alg:ctgan}. However, we skip the SGD updates on $p$ and instead sample directly from the GT $p$ and the uniform distribution and use~\eqref{eq:gen_loss} as the generator loss. 

\subsection{Experimental results}
\noindent \textbf{Quality of reconstructed image}: Fig.~\ref{fig:viz_results_clean}-\ref{fig:viz_results} compare the results of UVTomo-GAN jointly optimizing for $c$ and $p$ against the GT image and the aforementioned baselines for no noise and noisy scenarios. We also include the profiles of the middle vertical line of all methods against GT in Fig.~\ref{fig:line_profiles}. The results of UVTomo-GAN jointly optimizing for $c$ and $p$ closely resembles the oracle GAN-based given $p$ baseline, both qualitatively and quantitatively. However, with unknown $p$, the reconstruction problem is more challenging. Note that, although by assuming $p$ to be a uniform distribution (second to last column in Fig.~\ref{fig:viz_results_clean}-\ref{fig:viz_results}) the overall shape of the GT image emerges in the final reconstruction, the details are not successfully recovered. This highlights the importance of updating $p$ to retrieve details accurately in the reconstruction. A similar observation, although in a different setting is reported in~\cite{bora2018ambientgan, cryogan}. Furthermore, in the clean case, GLT is able to recover the correct ordering of the projection angles. However, as the projection angle distribution is non-uniform, assigning equi-spaced angles to the sorted projections causes a distorted reconstructed image (similar to GAN-based baseline with uniform $p$ assumption). On the other hand, in the MADE+GL baseline, as the projection angles are recovered, unlike projection sorting baseline i.e. GLT, it reconstructs the image accurately.

For $\textrm{SNR}=3$, while GLT's performance on the lung image is similar to the clean case, GLT's sorting of the projections for abdomen and ribosome images is erroneous despite tuning the hyperparameters (see Appendix~\ref{sec:baseline_detail}). Furthermore, we found the angle differences output by MADE for $\textrm{SNR}=3$ extremely noisy. This led to an erroneous angular difference estimation and incorrect projection embedding. As MADE+GL baseline failed in reconstructing all images in $\textrm{SNR}=3$, we excluded the results of this baseline in Fig.~\ref{fig:viz_results}.

In the presence of noise, we noticed that to obtain better results for EM starting from a random initialization, in the E-step~\eqref{eq:E-step}, we need to inflate the noise standard deviation $\sigma$, otherwise EM can get stuck easily in a poor local optima. In our EM experiments, we inflated $\sigma$ by a scalar factor of $\sqrt{2}$ for all datasets.

Fig.~\ref{fig:viz_results} also displays the effect of noise in the final reconstruction. We observe that the presence of noise makes the reconstruction task more challenging and degrades the reconstruction quality compared to the no noise case. This happens as the critic is having a harder time distinguishing signal from noise components given the noisy projections. 

\noindent \textbf{Quality of reconstructed $p$}: Comparison between the GT distribution of the projection angles and the one recovered by UVTomo-GAN with unknown $p$ is provided in Fig.~\ref{fig:pmf_results}. Note that the recovered $p$ matches the GT distribution both visually and quantitatively in terms of TV distance. Although, the quality of the recovered PMF in the noisy cases (Fig.~\ref{fig:pmf_results}-(b), (d), (e)) is not as good as the no noise case, it still closely resembles the GT projection distribution. This proves the ability of our approach to recover $p$ accurately under different distributions and noise regimes. 

\noindent \textbf{Convergence:}
We exhibit the convergence curves in terms of PSNR versus training iteration for no noise and noisy experiments in Fig.~\ref{fig:psnr_results}. To obtain this curve, at each iteration, we align the reconstructions with the GT. We compare the convergence of UVTomo-GAN for three cases, 1) given $p$, 2) assumed uniform $p$ and 3) jointly optimize $c$ and $p$. 

We noticed that once the dataset is augmented with flipped projections, for given $p$ setting, the convergence of the GAN training is more challenging. For a dataset with projection angles distributed in $[0, 2\pi)$, all possible rotations of the image, may constitute local optima of~\eqref{eq:minmax}. For given $p$, once the random initialization leads to a local optima, it is hard to get unstuck and recover the true distribution. 
On the other hand, for jointly optimizing $c$ and $p$ setting, as we have another degree of freedom $p$ to optimize, if the reconstructed image is rotated, the recovered $p$ will be accordingly rotated so that the synthetic distribution, matches the real one. 

Furthermore, when fixing the PMF with uniform distribution, after a certain number of iterations, we see no improvement in the reconstructed image. This is attributed to having an inaccurate PMF which hinders the correct distribution matching of synthetic and real measurements. Thus, the high frequency details in the final reconstructed image do not appear correctly (as also seen in Fig.~\ref{fig:viz_results_clean}-\ref{fig:viz_results}). This once again indicates the importance of recovering $p$ to have high quality reconstructions.

To evaluate the effect of using HB representation on the convergence, we compare against an experiment with pixel domain representation of the image. We call this baseline \textit{pixel UVTomo-GAN} versus our method \textit{HB UVTomo-GAN}. In this comparison, we use the same dataset, initialization, batch-size, learning rate decay and schedules for both pixel and HB UVTomo-GANs. For HB UVTomo-GAN, to only examine the effect of the representation, we use no TV regularization on the image, i.e. $\gamma_1 = 0$. However, for real UVTomo-GAN, to further help with the convergence, we set a small TV regularization weight as $5 \times 10^{-5}$ and enforce the image to be non-negative by defining it to be the output of a ReLU. For HB UVTomo-GAN, we choose $\alpha_\varphi = \alpha_c = 0.008$, $\alpha_p = 0.0008$ while for pixel UVTomo-GAN, we fine-tuned these parameters as $\alpha_\varphi = \alpha_I = 0.01$, $\alpha_p = 0.001$, $\alpha_I$ denoting the learning rate of the image. To implement the projection operator in pixel domain, we use Astra toolbox~\cite{aarle2016astra}.

In Fig.~\ref{fig:conv_viz_less_meas}, we show the results of this comparison. While both representations lead to accurate image and $p$ recovery, their convergence behaviours are different. For HB UVTomo-GAN, as we are operating in Hartley domain and the images tend to have larger low-frequency components compared to the high-frequency details, initially the gradients corresponding to lower frequency components are larger, leading to faster updates of $c_{k, q}$s for smaller $(k, q)$s. This helps in more stable convergence of HB versus pixel UVTomo-GAN.

Note that, for HB UVTomo-GAN, we obtain a reasonable image and PMF at early stages of training, i.e., after $20$k-$40$k iterations (which takes roughly $6$-$12$ minutes). As expected, the image is further refined with more training iterations.

\vspace{-0.15cm}
\section{Conclusion}
\label{sec:conclusion}
In this paper, we proposed an adversarial learning approach for the unknown view tomographic reconstruction problem. We presumed the projection angles and the probability distribution they are drawn from are not known a-priori. Thus, we recovered both the unknown image and probability distribution of the projection angles via a distribution matching formulation solved through a min-max game between a critic and a generator. To further reduce computational burdens, we employed a Fourier related representation of the image, expanded on a Hartley-Bessel basis set. For the GAN training, we showed that the loss function at the generator side is non-differentiable with respect to the projection angle distribution. Thus, we used the Gubmel-Softmax approximation of samples from discrete distributions. We studied the theoretical guarantees of UVTomo-GAN and demonstrated that asymptotically unique recovery of the image and projection distribution is achieved. Our simulation results confirmed the capability of our method in accurate image and projection angle distribution recovery under different noise regimes.

\vspace{-0.2cm}
\section{Appendix}
\label{sec:appendix}
\subsection{Computational cost of UVTomo-GAN}
\label{sec:cost_app}
\noindent \textbf{Cost of projection generation}: To generate $N_\theta=O(m)$ projection templates following~\eqref{eq:hartley_projection2}, we first compute the inner summation over $q$, i.e.,
\begin{align}
   f_k(\xi_j) = \sum\limits_{q=1}^{p_k} c_{k, q} J_{k, q}(\xi_j).
\end{align}
On the radial line, we have $O(m)$ equally spaced points $\xi_j$. Given that $K_{\textrm{max}}=O(m)$ and $p_k=O(m)$, computing $f_k(\xi_j)$, $\forall k, j$ requires $O(m^3)$ computations. 

Next, using $f_k(\xi_j)$ we compute the outer sum in~\eqref{eq:hartley_projection2} with respect to $k$ for $N_\theta$ projection angles. A naive matrix multiplication implementation for this step leads to $O(m^3)$ cost (multiplying two matrices of size $O(m)\times O(m)$). This can be further reduced using FFT to $O(m^2 \log m)$. Finally, the total cost of generating $N_{\theta}$ projections using~\eqref{eq:hartley_projection2} is $O(m^3)$. 

\vspace{-0.1cm}
\subsection{Proof of Theorem 1}
\label{sec:proof_thm1}
First we prove:
\begin{align}
    P^{\textrm{clean}}_{f, p_f} = P^{\textrm{clean}}_{g, p_g} \Rightarrow [f] = [g].
\end{align}
From $P^{\textrm{clean}}_{f, p_f} = P^{\textrm{clean}}_{g, p_g}$, it is implied that the support of the two distributions are the same. This means that $f$ and $g$ have the same projection set. In other words, $\{\mathcal{P}_{\theta_i} f\}_{i=1}^{N_{\theta}} = \{\mathcal{P}_{\widehat{\theta}_j} g\}_{j=1}^{N_{\theta}}$ where $\widehat{\bm{\theta}} = \{\widehat{\theta}_j\}_{j=1}^{N_\theta}$ can be a shuffled version of $\bm{\theta} = \{{\theta}_i\}_{i=1}^{N_\theta}$. Intuitively, one can imagine two objects $f$ and $g$ which have the same projections, however the order of the projection angles of $f$ can be a shuffled version of the projection angles for $g$. Now the question that arises is: Given the class of functions $f$ and $g$ belong to, is it possible to have two distinct objects that produce identical projection sets?

This question is related to the feasibility of unique angle recovery in unknown view tomography, comprehensively studied in~\cite{Basu2000, Basu2000_1}. Based on our discussions so far, we seek to prove the following:
\begin{align}
    \{\mathcal{P}_{\theta_i} f\}_{i=1}^{N_{\theta}} = \{\mathcal{P}_{\widehat{\theta}_j} g\}_{j=1}^{N_{\theta}} \Rightarrow [f]=[g]. \label{eq:proj_set_eq}
\end{align}
In~\eqref{eq:proj_set_eq}, the LHS implies that $f$ and $g$ have the same set of projections, in other words we have: $\forall \gamma \in \{\mathcal{P}_{\theta_i} f\}_{i=1}^{N_{\theta}}$, $\gamma \in \{\mathcal{P}_{\widehat{\theta}_j} g\}_{j=1}^{N_{\theta}}$ and $\forall \gamma' \in \{\mathcal{P}_{\widehat{\theta}_j} g\}_{j=1}^{N_{\theta}}$, $\gamma' \in \{\mathcal{P}_{\theta_i} f\}_{i=1}^{N_{\theta}}$. To prove the above, we borrow the definitions and various theoretical results in~\cite{Basu2000}. Helgasson–Ludwig (HL) consistency conditions~\cite{Natterer2001} link the geometric moments of a 2D object to its projections. Let $v$ and $\mu$ define the geometric moment of the image $f$ and its projection as:
\begin{align}
    v_{i, k}(f) &= \int\limits_{-1}^{1} \int\limits_{-1}^{1} x^i y^k f(x, y) dx dy \\
    \mu_d(\theta; f) &= \int\limits_{-1}^{1} x^d \{\mathcal{P}_{\theta} f\}(x) dx.
\end{align}
Object moments of order $d$ are the ones that satisfy $i+k=d$. Let $\mathbf{v}(f)$, denote the set of geometric moments of order $d \in D$ for object $f$. Given the object moments $\mathbf{v}(f)$, we construct a family of trigonometric polynomials as:
\begin{align}
    \mathcal{Q}_d(\theta; \mathbf{v}(f)) = \sum\limits_{r=0}^d {d \choose r} v_{r, d-r}(f) \, (\cos \theta)^r (\sin \theta)^{d-r}. \label{eq:Q_def} 
\end{align}
Given the definition~\eqref{eq:Q_def}, we state the HL conditions as:
\begin{align}
    \mathcal{Q}_d(\theta ; \mathbf{v}(f)) = \mu_d(\theta; f). \label{eq:hl}
\end{align}
We have defined equivalence for 2D images before. If two images are equivalent, then they are related through a rotation and reflection. Similarly, we can define equivalence on the projection angles. Assume two vector of projection angles of length $N_\theta$, $\bm{\theta} , \widehat{\bm{\theta}} \in [-\pi, \pi]^{N_\theta}$. $\bm{\theta}$ is said to be equivalent to $\widehat{\bm{\theta}}$, i.e., $\bm{\theta} \sim \widehat{\bm{\theta}}$ if $\exists \eta \in \{-1, 1\}$ and $\alpha \in [-\pi, \pi]$ such that $\widehat{\theta}_i = \eta \theta_i + \alpha + 2 \pi n_i$, for $n_i \in \mathbb{Z}$.

As the projection set for $f$ and $g$ objects are the same (based on~\eqref{eq:proj_set_eq}), we conclude $\forall \theta$, $\exists \widehat{\theta}$ such that:
\begin{align}
    \mu_d(\theta; f) = \mu_d(\widehat{\theta}; g). \label{eq:proj_moments}
\end{align}
After invoking HL conditions~\eqref{eq:hl} for object $f$ on the RHS of~\eqref{eq:proj_moments} we get:
\begin{align}
    \mathcal{Q}_d(\theta ; \mathbf{v}(f)) = \mu_d(\widehat{\theta};g), \quad \forall d\geq0. \label{eq:arp}
\end{align}
Note that, we have narrowed down the identical projection sets for $f$ and $g$ to~\eqref{eq:arp}. Now we restate our question as: what is the relationship between $\theta$ and $\widehat{\theta}$?

To find the answer to this question, we first limit the set of moment orders to $d \in D=\{1, 2\}$ (as~\eqref{eq:arp} holds for $\forall d\geq0$, we can simply do this). Note that, for $\theta\in[0, 2\pi)$, the projections corresponding to $\theta \in [\pi, 2\pi)$ are a flipped version of projections associated to $\theta \in [0, \pi)$ and do not constitute new information~\cite{Basu2000}. Thus, in~\cite{Basu2000}, the authors limit their analysis to the projections that are $\pi$-distinct, i.e. there are no two angles that are different by a factor of $\pi$. Following the same lines, given the projection sets corresponding to $\theta, \widehat{\theta} \in [0, 2\pi)$, we select a $\pi$-distinct projection subset by choosing a set of projections that have positive (or negative) 1st order geometric moment. We now invoke Corollary 5 of Theorem 9 in~\cite{Basu2000}. We restate this corollary in the following.

\noindent \textbf{Corollary 1 }(Corollary 5 of Theorem 9~\cite{Basu2000}): Suppose $\theta$ is a set of $\pi$-distinct view angles and $N_\theta>8$. Suppose $\mathbf{v}$ satisfies the following condition: $\not \exists \beta, \gamma \in \mathbb{R}$ such that:
\begin{align}
    \mathcal{Q}_2(\theta; \mathbf{v}) = \beta \left( \mathcal{Q}_1(\theta; \mathbf{v}) \right)^2 + \gamma, \, \forall \theta \in [0, \pi] \label{eq:condition_m}
\end{align}
or equivalently,
\begin{align}
    \textrm{det}\left[\begin{array}{ccc}
    v_{1, 0}^2 & v_{2, 0} & 1\\
    2 v_{1, 0} v_{0, 1} & v_{1, 1} & 0\\
    v_{0, 1}^2 & v_{0, 2} & 1
    \end{array}
    \right]\neq 0. \label{eq:condition_det}
\end{align}
If $\theta \not \in \textrm{UAS}(\mathbf{v})$ with $\textrm{UAS}$ (unidentifiable angle set) defined as:
\begin{align}
    \textrm{UAS}(\mathbf{v}) = \left\{ \arg\left(\sqrt{\frac{-c^*_1}{c_1}}\right), \arg\left(-\sqrt{\frac{-c^*_1}{c_1}}\right) \right\}
\end{align}
where,
\begin{align}
    c_1 = \frac{1}{2}(v_{1, 0}- i \, v_{0, 1})
\end{align}
then, the only view angles $\widehat{\theta}$ that produce the same projection moments of order $D=\{1, 2\}$ are equivalent to $\theta$. This implies that $\theta \sim \widehat{\theta}$. $\square$

Adhering to Corollary 1, if $\mathbf{v}(f)$ satisfies the conditions in~\eqref{eq:condition_m} or~\eqref{eq:condition_det}, then for $d \in \{1, 2\}$, the only projection angles $\widehat{\theta}$ for which~\eqref{eq:arp} holds are equivalent to $\theta$ and thus $\theta \sim \widehat{\theta}$. On the other hand, based on Corollary 1, the projection angles recovered for $f$, i.e., $\theta$ are equivalent to the GT projection angles $\check{\theta}$ used for generating the projections of $f$, i.e, $\theta \sim \check{\theta}$. Based on the transitivity property of equivalence relation, this leads to $\widehat{\theta} \sim \check{\theta}$

Given $\theta \sim \widehat{\theta} \sim \check{\theta}$ and the fact that the projection sets corresponding to the objects $f$ and $g$ are identical, the objects $\widehat{f}$ and $\widehat{g}$ reconstructed from the projection sets and projection angles would also be the same (up to a rotation and reflection), i.e., $[\widehat{f}] = [\widehat{g}]$. We now link the reconstructed objects and their ground truths. 

If we have sufficiently large $N_\theta$, we can directly recover HB expansion coefficients $c$ by solving a set of linear equations linking the projections to the HB expansion coefficients. Given the HB expansion coefficients, we have a continuous representation of the image as defined in~\eqref{eq:HB_spatial}. This leads to $\widehat{f} = f$ and $\widehat{g} = g$ and finally concludes $[f] = [g]$. 

As $[f] \!=\! [g]$, $\exists \, \Gamma \! \in \! \mathrm{O}(2)$ such that $g \!=\! \Gamma f$. $P^{\textrm{clean}}_{f, p_f} \! =\! P^{\textrm{clean}}_{g, p_g}$ implies the TV distance between the two probability distributions is zero, i.e., 
\begin{align}
    TV(P^{\textrm{clean}}_{f, p_f}, P^{\textrm{clean}}_{\Gamma f, p_g}) = 0. \label{eq:clean_dist}
\end{align}
Invoking Lemma 1 (stated in Appendix~\ref{ssec:lemma1}), we know $P^{\textrm{clean}}_{\Gamma f, p_g} = P^{\textrm{clean}}_{f, \Gamma^{-1} p_g}$, therefore~\eqref{eq:clean_dist} becomes,
\begin{align}
    TV(P^{\textrm{clean}}_{f, p_f}, P^{\textrm{clean}}_{f, \Gamma^{-1} p_g}) &= TV(p_f, \Gamma^{-1} p_g) \nonumber \\
    &= \frac{1}{2} \Vert p_f -  \Gamma^{-1} p_g \Vert_1. \label{eq:tv} 
\end{align}
Following~\eqref{eq:clean_dist}, the LHS of~\eqref{eq:tv} is $0$. Thus, based on the non-negativity property of $\Vert . \Vert_1$ norm, we have,
\begin{align}
    p_f = \Gamma^{-1} p_g \Rightarrow p_g = \Gamma p_f
\end{align}
implying $[p_f] = [p_g]$. $\blacksquare$

\subsection{Proof of Theorem 2}
\label{sec:proof_thm2}

Our proof follows closely the proof of Theorem 1 in~\cite{cryogan}. We first show that,
\begin{align}
    P^{\textrm{noisy}}_{f, p} = P^{\textrm{noisy}}_{\widetilde{f}, \widetilde{p}} \Rightarrow P^{\textrm{clean}}_{f, p} = P^{\textrm{clean}}_{\widetilde{f}, \widetilde{p}}. \label{eq:noisy_clean}
\end{align}
According to the forward model~\eqref{eq:proj_noisy}, we have $\zeta = \mathcal{P}_{\theta} f + \varepsilon$ where $\varepsilon[n] \sim q_{\epsilon}$ an IID additive noise which is independent of $f$ and $\theta \sim p$. Note that we are considering a general model for the noise and not confining it to be a Gaussian. As $\varepsilon$ is independent of the image and projection angles, we have:
\begin{align}
    P^{\textrm{noisy}}_{f, p} = P^{\textrm{clean}}_{f,p} * q_{\varepsilon} \label{eq:conv_pdf}
\end{align}
In Fourier domain,~\eqref{eq:conv_pdf} becomes:
\begin{align}
    \mathcal{F} \{P^{\textrm{noisy}}_{f, p}\} = \mathcal{F} \{P^{\textrm{clean}}_{f,p}\} \mathcal{F} \{q_{\varepsilon}\}. \label{eq:conv_pdf_fd}
\end{align}
We have assumed $\varepsilon$ to have full support in Fourier domain, therefore we can divide both sides of~\eqref{eq:conv_pdf_fd} by $\mathcal{F} \{p_{\varepsilon}\}$. Therefore given $\mathcal{F} \{P^{\textrm{noisy}}_{f, p}\}$, we have $\mathcal{F} \{P^{\textrm{clean}}_{f, p}\}$ and~\eqref{eq:noisy_clean} is proved. Now, we show:
\begin{align}
    P^{\textrm{clean}}_{f, p} = P^{\textrm{clean}}_{\widetilde{f}, \widetilde{p}} \Rightarrow \widetilde{f} = \Gamma f \textrm{ and } \widetilde{p} = \Gamma p \label{eq:transform_res}
\end{align}
for a unique $\Gamma \in \mathrm{O}(2)$. To prove~\eqref{eq:transform_res}, we invoke Theorem 1. Theorem 1 states that if the two images $f$ and $\widetilde{f}$ have the same distribution of the clean projections, then the objects and their associated projection distributions are equivalent up to a rotation and reflection. This confirms $[f] = [\widetilde{f}]$, and $[p] = [\widetilde{p}]$, i.e., $\widetilde{f} = \Gamma f$ and $\widetilde{p} = \Gamma p$, for a $\Gamma \in \mathrm{O}(2)$.$\blacksquare$ 

\vspace{-0.5cm}
\subsection{Lemma 1}
\label{ssec:lemma1}
Assume $f \! \in \! \mathcal{L}_1(\mathbb{B}_2) \cap \mathcal{L}_2(\mathbb{B}_2)$, projection angles $\theta$ are distributed following $p$, i.e. $\theta \sim p$ and $\Gamma \in \mathrm{O}(2)$. Then, 
\begin{align}
    P^{\textrm{clean}}_{f, \Gamma^{-1} p} = P^{\textrm{clean}}_{\Gamma f, p}
\end{align}
\textit{Proof:} For a given ($f$, $p_f$), if $\gamma \in \mathrm{O}(2)$ is applied to both $f$ and $p$, then the induced probability distribution of the projection images would be the same, i.e. $P^{\textrm{clean}}_{f, p} = P^{\textrm{clean}}_{\Gamma f, \Gamma p}$. After changing $p' = \Gamma p$, we have $P^{\textrm{clean}}_{f, \Gamma^{-1} p'} = P^{\textrm{clean}}_{\Gamma f, p'}$, thus concluding the proof.

\vspace{-0.5cm}
\subsection{Details on baselines}
\label{sec:baseline_detail}

\textbf{GLT~\cite{Singer2013}: } For this baseline, a graph is constructed based on the pairwise distances of the compressed denoised projections. The tunable parameters in GLT are 1) number of nearest neighbors (NN) for each projection, 2) Jaccard index threshold ($\beta$). The choice of NN affects the connectivity of the constructed graph (before denoising). On the other hand, Jaccard index thresholding reduces the shortcut edges in the graph. For the clean case, we choose $\textrm{NN} = 111$ and $\beta = 0.41$. In the noisy case, we set $\textrm{NN}=111$ and $\beta = 0.21$, $\beta=0.31$ and $\beta=0.41$ for Lung, Abdomen and Ribosome images, respectively.

\textbf{MADE~\cite{Phan2017} + GL: } To find the angular differences between any two projections we use MADE. The tunable parameters for MADE are similar to GLT. For the lung and abdomen images, we set the number of nearest neighbors $\textrm{NN}=70$ while $\textrm{NN}=90$ for the ribosome image. For all the images, we set $\beta = 0.1$. After obtaining, the angular differences between the neighborhood projections, through a shortest path algorithm, i.e. Djikstra, the absolute angle differences between any two projections are obtained. Next, we construct a weight matrix $E$ based on the angle differences from MADE as:
\begin{align}
    E(i, j) = 
    \begin{cases}
    e^{- \frac{\vert \theta_i - \theta_j \vert^2}{\epsilon}}, \quad &\vert \theta_i - \theta_j \vert \leq 5^\circ \\
    0. \quad &\textrm{o.w.}
    \end{cases}
\end{align}
where $\theta_i$ denotes the angle corresponding to the $i$-the projection. In our experiments, we set $\epsilon = 20$. Next, we normalize $E$ similar to~\cite{Singer2013} and perform eigenvalue decomposition. In the clean case, the top two non-trivial eigenvectors of the normalized matrix form the embedding of the projections which is a circle. The angle of the $i$-th projection embedded on the circle is assigned as $\theta_i$. Based on the assigned projection angles, the image is reconstructed.

\section{Acknowledgement}
The authors would like to thank Prof. Amit Singer and Prof. Hau-Tieng Wu for their helpful suggestions and sharing the code used in one of our baselines.
\small{
\bibliographystyle{ieeetr}
\bibliography{refs.bib}}

\end{document}